\documentclass[conference]{IEEEtran}
\IEEEoverridecommandlockouts
\usepackage{cite}
\usepackage{amsmath,amssymb,amsfonts}
\usepackage{graphicx}
\usepackage{textcomp}
\usepackage{xcolor}
\def\BibTeX{{\rm B\kern-.05em{\sc i\kern-.025em b}\kern-.08em
    T\kern-.1667em\lower.7ex\hbox{E}\kern-.125emX}}

\newenvironment{customlegend}[1][]{%
  \begingroup
  \csname pgfplots@init@cleared@structures\endcsname
  \pgfplotsset{#1}%
}{%
  \csname pgfplots@createlegend\endcsname
  \endgroup
}%
\def\addlegendimage{\csname pgfplots@addlegendimage\endcsname}
\usepackage{float}
\usepackage{subcaption}
\usepackage{tikz}
\usetikzlibrary{patterns}
\usepackage{pgfplots}
\usepackage{filecontents}
\usepackage{bbding}
\usepackage{threeparttable}
\usepackage{enumitem}
\usepackage{ulem}
\usepackage{balance}

\makeatletter
\newif\if@restonecol
\makeatother

\usepackage[linesnumbered,ruled,vlined]{algorithm2e}
\usepackage{algpseudocode}
\usepackage{amsmath}
\usepackage{booktabs}

\usepackage{url}
\usepackage{xspace}
\usepackage{multirow}
\usepackage{graphicx}
\usepackage{colortbl}
\newcommand{\ie}{{\it i.e.},\xspace}
\newcommand{\eg}{{\it e.g.},\xspace}

\newcommand{\etalstd}{{et~al.}\xspace}

\def\header{\vspace{0.6mm} \noindent}

\newcommand{\fig}{{Fig.}\xspace}
\newcommand{\tab}{{Table}\xspace}

\newcommand{\algo}{\textsf{{\small TRMMA}}\xspace}
\newcommand{\rec}{\textsf{{\small TRMMA}}\xspace}
\newcommand{\gts}{\textsf{{\small MMA}}\xspace}
\newcommand{\Enc}{{DualFormer Encoding}\xspace}

\newcommand{\enc}{{DualFormer encoding module}\xspace}
\newcommand{\Dec}{{Decoding}\xspace}
\newcommand{\dec}{{decoding module}\xspace}

\newcommand{\nn}{\textsf{{\small Nearest}}\xspace}
\newcommand{\nnlinear}{\textsf{{\small Nearest+linear}}\xspace}

\newcommand{\MMAlinear}{\textsf{{\small MMA+linear}}\xspace}
\newcommand{\fmm}{\textsf{{\small FMM}}\xspace}
\newcommand{\graphmm}{\textsf{{\small 
 GraphMM}}\xspace}
\newcommand{\deepmm}{\textsf{{\small DeepMM}}\xspace}
\newcommand{\lhmm}{\textsf{{\small LHMM}}\xspace}

\newcommand{\linear}{\textsf{{\small Linear}}\xspace}
\newcommand{\dhtr}{\textsf{{\small DHTR}}\xspace}
\newcommand{\mtrajrec}{\textsf{{\small MTrajRec}}\xspace}
\newcommand{\rntrajrec}{\textsf{{\small RNTrajRec}}\xspace}
\newcommand{\teri}{\textsf{{\small TERI}}\xspace}
\newcommand{\stvec}{\textsf{{\small ST2Vec+Dec}}\xspace}
\newcommand{\trajgat}{\textsf{{\small TrajGAT+Dec}}\xspace}
\newcommand{\trajcl}{\textsf{{\small TrajCL+Dec}}\xspace}

\newcommand{\stged}{\textsf{{\small MM-STGED}}\xspace}

\newcommand{\sprime}{^{\prime}}

\newcommand{\samp}{\gamma\xspace}

\newcommand{\eps}{\epsilon\xspace}
\newcommand{\pgps}{p\xspace}
\newcommand{\pseg}{a\xspace}
\newcommand{\pspred}{\hat{\pseg}\xspace}
\newcommand{\pii}{\pgps_i\xspace}
\newcommand{\piemb}{\mathbf{\pgps}_i\xspace}
\newcommand{\pis}{\pseg_i\xspace}
\newcommand{\pjs}{\pseg_j\xspace}
\newcommand{\ps}[1]{\pseg_{#1}\xspace}
\newcommand{\zia}{\mathbf{z}_i^{(0)}\xspace}
\newcommand{\zib}{\mathbf{z}_i^{(1)}\xspace}
\newcommand{\zb}[1]{\mathbf{z}_{#1}^{(1)}\xspace}
\newcommand{\Zb}{\mathbf{Z}_1\xspace}
\newcommand{\zic}{\mathbf{z}_i^{(2)}\xspace}
\newcommand{\zc}[1]{\mathbf{z}_{#1}^{(2)}\xspace}
\newcommand{\Zc}{\mathbf{Z}_2\xspace}

\newcommand{\cj}{c_j\xspace}
\newcommand{\cjemb}{\mathbf{c}_j\xspace}

\newcommand{\tgps}{T\xspace}
\newcommand{\trajhigh}{\mathcal{T}_\epsilon\xspace}
\newcommand{\thighpred}{\hat{\mathcal{T}}_\epsilon\xspace}
\newcommand{\thigh}{\trajhigh}

\newcommand{\TM}{\mathbf{T}\xspace}
\newcommand{\Ta}{\mathbf{T}_0\xspace}
\newcommand{\Tb}{\mathbf{T}_1\xspace}

\newcommand{\RM}{\mathbf{R}\xspace}

\newcommand{\Rb}{\mathbf{R}_1\xspace}

\newcommand{\lt}{\ell\xspace}
\newcommand{\lR}{\ell_{\Route}\xspace}
\newcommand{\lthigh}{\ell_\eps\xspace}
\newcommand{\HM}{\mathbf{H}\xspace}

\newcommand{\hj}{\mathbf{h}_j\xspace}
\newcommand{\hv}{\mathbf{h}\xspace}

\newcommand{\numseg}{n\xspace}
\newcommand{\numinter}{m\xspace}

\newcommand{\rnet}{G\xspace}

\newcommand{\routepred}{\mathcal{R}\xspace}

\newcommand{\Route}{\routepred}
\newcommand{\trajdata}{\mathcal{D}\xspace}
\newcommand{\onehot}{\mathbf{1}\xspace}
\newcommand{\nvmatorigin}{\mathbf{W}_{G}\xspace}

\newcommand{\keyprob}{P\xspace}

\newcommand{\candset}{\mathcal{C}}
\newcommand{\candsize}{k_c}
\newcommand{\Cpi}{\candset_{\pgps_i}\xspace}

\definecolor{forestgreen}{RGB}{34, 139, 34}
\definecolor{RYB1}{RGB}{192, 128, 255}
\definecolor{RYB2}{RGB}{255, 192, 32}
\definecolor{RYB3}{RGB}{139, 0, 0}
\definecolor{RYB4}{RGB}{0, 128, 255}

\newcommand{\first}{\textbf}
\newcommand{\second}{\underline} 
\newcommand{\third}{\underline}

\definecolor{myred}{HTML}{fd7f6f}
\definecolor{mywhite}{HTML}{D8D8D8}
\definecolor{myorange}{HTML}{D7191C}
\definecolor{myblue}{HTML}{7eb0d5}
\definecolor{mygreen}{HTML}{b2e061}
\definecolor{mypurple}{HTML}{bd7ebe}
\definecolor{myorange}{HTML}{ffb55a}
\definecolor{myyellow}{HTML}{ffee65}
\definecolor{mypurple2}{HTML}{beb9db}
\definecolor{mypink}{HTML}{fdcce5}
\definecolor{mycyan}{HTML}{8bd3c7}
\definecolor{mycyan2}{HTML}{00ffff}
\definecolor{myblue2}{HTML}{115f9a}
\definecolor{myred2}{HTML}{c23728}

\newtheorem{definition}{Definition}

\pgfplotsset{compat=1.18}
\begin{document}

\title{
Efficient Methods for Accurate Sparse Trajectory Recovery and Map Matching
}

\author{\IEEEauthorblockN{Wei Tian}
\IEEEauthorblockA{\textit{Department of Computing} \\
\textit{Hong Kong Polytechnic University} \\
wei.tian@connect.polyu.hk}
\and
\IEEEauthorblockN{Jieming Shi$^*$\thanks{*Corresponding author.}}
\IEEEauthorblockA{\textit{Department of Computing} \\
\textit{Smart Cities Research Institute} \\
\textit{Hong Kong Polytechnic University}\\
jieming.shi@polyu.edu.hk}
\and
\IEEEauthorblockN{Man Lung Yiu}
\IEEEauthorblockA{\textit{Department of Computing} \\
\textit{Hong Kong Polytechnic University} \\
csmlyiu@comp.polyu.edu.hk}
}

\maketitle

\begin{abstract}

Real-world trajectories are often sparse with low-sampling rates (i.e., long intervals between consecutive GPS points) and misaligned with road networks, yet many applications demand high-quality data for optimal performance. To improve data quality with sparse trajectories as input, we systematically study two related research problems: \textit{trajectory recovery on road network}, which aims to infer missing points to recover high-sampling trajectories, and \textit{map matching}, which aims to map GPS points to road segments to determine underlying routes. Capturing latent patterns in complex sparse trajectory data on road networks is challenging, especially with large-scale datasets.

In this paper, we present efficient methods \algo and \gts for accurate trajectory recovery and map matching, respectively, where \gts serves as the first step of \algo. In \gts, we carefully formulate a classification task to map a GPS point from sparse trajectories to a road segment over a small candidate segment set, rather than the entire road network. We develop techniques in \gts to generate effective embeddings that capture the patterns of GPS data, directional information, and road segments, to accurately align sparse trajectories to routes. For trajectory recovery, \algo focuses on the segments in the route returned by \gts to infer missing points with position ratios on road segments, producing high-sampling trajectories efficiently by avoiding evaluation of all road segments. Specifically, in \algo, we design a dual-transformer encoding process to cohesively capture latent patterns in trajectories and routes, and an effective decoding technique to sequentially predict the position ratios and road segments of missing points.
We conduct extensive experiments to compare \algo and \gts with numerous existing methods for trajectory recovery and map matching, respectively, on 4 large  real-world datasets.
\algo and \gts consistently achieve the best result quality, often by a significant margin.
Moreover, \algo and \gts are highly efficient during training and inference, being up orders of magnitude faster  than the next best competitors. 
The implementation is at \url{https://github.com/derekwtian/TRMMA}.

\end{abstract}


\section{Introduction}\label{sec:intro}
GPS trajectories, \ie sequences of timestamped GPS points, are increasingly produced on road networks due to the widespread use of mobile devices. High-quality trajectory data are crucial for many applications, including vehicle navigation~\cite{NMLR21,DRPK23}, travel time estimation~\cite{DeepOD20,Yuan00020B22}, trajectory similarity search~\cite{VRE2022,FangG0XGJ23,Jin000023}, and traffic flow predication~\cite{FangPCDG21,YuanCL24}. 

Ideally, trajectories should have high sampling rates with short intervals between GPS points (e.g., every 15 seconds)~\cite{RNTrajRec23} and be aligned with road segments~\cite{LHMM24ICDE}.
However, real trajectories are often sparse and of low quality, characterized by (i) low sampling rates with long intervals between consecutive GPS points (\eg every 2–6 minutes \cite{mm10,RNTrajRec23}), and (ii) inaccurate GPS coordinates deviating from road segments due to device or signal limits~\cite{GraphMM24TKDE,LHMM24ICDE}. Sparse trajectories lose significant travel information on road networks, negatively impacting  the aforementioned applications.

\begin{figure}[!t]
\centering
\begin{small}
\includegraphics[width=0.92\columnwidth]{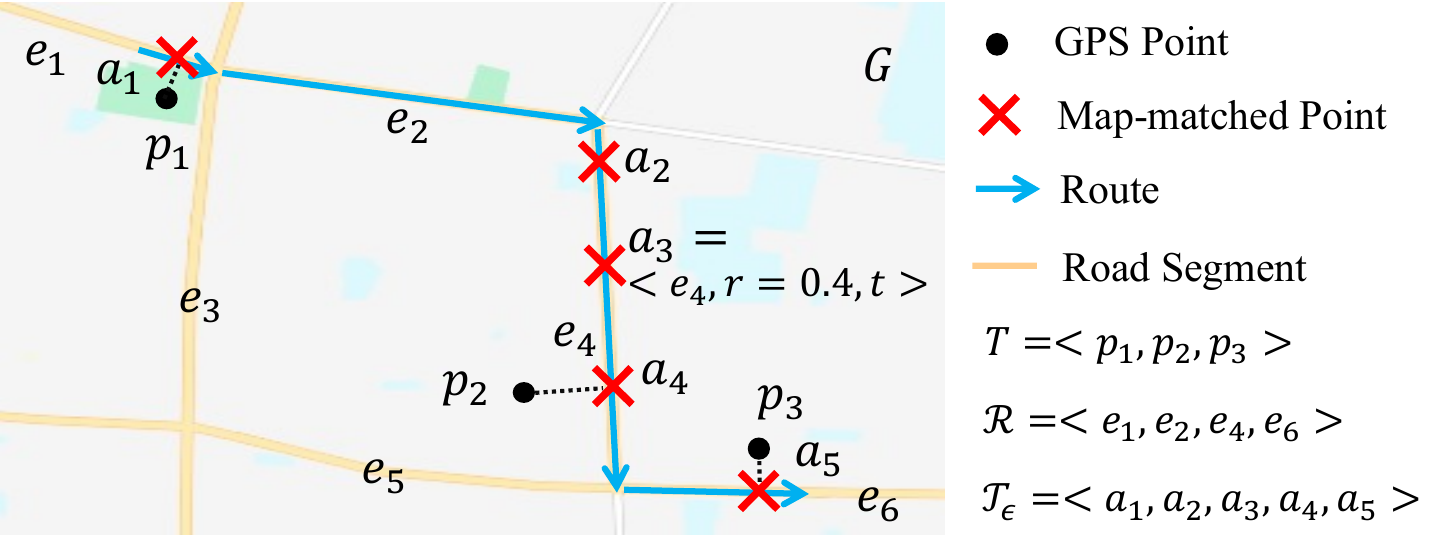}
\vspace{-0mm}
\caption{Example of Trajectory Recovery on Road Network and Map Matching}\label{fig:eg-rnet}
\vspace{-3mm}
\end{small}
\end{figure}

To mitigate the issues of sparse trajectories and improve data quality, we study two closely related research problems in a principled way: \textit{trajectory recovery on road network}~\cite{RNTrajRec23,MTrajRec21} and \textit{map matching}~\cite{GraphMM24TKDE}.
{Briefly, map matching seeks to identify the segments of GPS points and find the route of a trajectory, while trajectory recovery is to infer the missing points of a sparse trajectory to achieve a target high-sampling rate. 
These two tasks enhance data quality and are important in applications such as navigation~\cite{ShiTZ0WY24}, travel time estimation~\cite{Yuan00020B22}, and traffic flow analysis~\cite{FangPCDG21}.} 
We use \fig~\ref{fig:eg-rnet} to illustrate the two tasks. 
In the  road network $\rnet$ shown in \fig~\ref{fig:eg-rnet}, the three GPS points  $\pgps_1,\pgps_2,\pgps_3$ (black dots) form a sparse trajectory $\tgps$ with a low-sampling rate. Observe that the GPS points are not on any segment. Map matching aims to identify $\tgps$'s underlying route $\Route$, a sequence of road segments $e_1,e_2,e_4,e_6$ (blue arrows). 
Given a target high-sampling rate $\eps$, trajectory recovery on road network is to infer the missing points of sparse trajectory $\tgps$. The inferred points $\pseg_1,\pseg_2,\pseg_3,\pseg_4,\pseg_5$ (red crosses) are aligned onto road segments with position ratios indicating their relative positions. These inferred points form a high-sampling trajectory $\thigh$ for $\tgps$. 
Please see Section~\ref{sec:preliminary} for detailed definitions and discussion.

Effective map-matching and trajectory recovery are both challenging, {especially with large datasets containing millions of trajectories on road networks.
Capturing the latent patterns in GPS points, trajectory sequences, and road segments is non-trivial and computationally expensive.}
Early map matching methods~\cite{PinkH08,HMM09} mainly adopt Hidden Markov Model (HMM) with heuristics to identify the route of a GPS trajectory, which suffer from degraded performance, {especially over sparse GPS trajectories~\cite{HMM09}}.
Recent map matching studies~\cite{DMM24TKDE,GraphMM24TKDE,LHMM24ICDE} develop end-to-end methods incorporating learning techniques to extract patterns from historical trajectories. Nevertheless, the map matching methods can only find routes but not  increase the sampling rates of  trajectories.

As for trajectory recovery on road network, it has recently attracted research attention~\cite{MTrajRec21,RNTrajRec23,STGED24}.
For instance, 
RNTrajRec~\cite{RNTrajRec23} considers road network structure and the surrounding subgraphs of observed GPS points, utilizing grids and graph neural networks, for trajectory recovery.
However, these methods~\cite{RNTrajRec23,MTrajRec21} consider all segments in a road network as candidates to infer missing points, leading to suboptimal performance and intensive computational overhead. 
Other approaches~\cite{DHTR21,TERI23,MuslehM23} work on trajectory recovery in free space as a grid and tend to underperform when adapted to road networks.
Despite the close relationship between trajectory recovery and map matching, existing studies are often dedicated to one problem.

Given a sparse trajectory $\tgps$, a natural methodology for trajectory recovery on a road network $\rnet$ is to first use map matching as a  procedure to align the GPS points in $\tgps$ onto road segments in $\rnet$, obtaining route $\Route$, and subsequently recover the missing points over $\Route$. 
The literature contains rough discussions on this idea but predominantly emphasizes its ineffectiveness~\cite{STGED24,RNTrajRec23}, caused by the severe sparsity of input raw GPS trajectories.
However, we believe this critique warrants further thorough investigation.

Hence, in this paper, we solve the two problems together, fulfilling the potential of the promising but undervalued idea above, and present efficient and accurate methods: \algo for \underline{T}rajectory \underline{R}ecovery on road networks, with \gts for \underline{M}ap \underline{MA}tching as a procedure invoked in \algo.

For a sparse trajectory $\tgps$, \algo starts by invoking \gts to map the GPS points in $\tgps$ to their segments in a road network $\rnet$. The entire $\rnet$ may contain numerous segments and trajectory $\tgps$ is sparse,  posing great challenges for \gts to be accurate and fast. To address these challenges, in \gts, we conduct empirical analysis and formulate the task of identifying the segment of a GPS point $\pii$ as a classification problem over a small candidate segment set, rather than the entire $\rnet$. We then design techniques to produce  effective embeddings for $\pii$ and its candidate segments by holistically considering GPS data, directional information, road network context, and the sequential patterns of $\tgps$. The embeddings are used to map a GPS point to a segment in \gts. Over the predicted segments of the points in $\tgps$, \gts deduces the underlying route $\Route$.
\gts achieves high accuracy in real-world datasets, serving as a solid basis for \algo to conduct trajectory recovery.  
Specifically, \algo restores the missing points along the route $\Route$ of $\tgps$ to produce a high-sampling $\thigh$ that satisfies a specified sampling rate. Note that \rec concentrates solely on the segments in $\Route$ to infer missing points, avoiding the expensive evaluation of all segments within road network $\rnet$, thereby improving efficiency. \rec requires only the sparse trajectory $\tgps$ and its route $\Route$ as inputs. In \rec, we design a dual-transformer encoding module to produce effective embeddings that capture the patterns in $\tgps$ and $\Route$, synergized via an attention mechanism. Leveraging the embeddings, \rec uses a multitask decoding process to sequentially predict missing points, with classification over the candidate segments in $\Route$ and regression to estimate position ratios on these segments. 
We conduct comprehensive experiments across large-scale trajectory datasets on road networks, showing that \algo and \gts consistently surpass existing methods for trajectory recovery and map matching, respectively, in terms of result quality and efficiency. For instance, on a large PT dataset, 
\algo only needs 0.88 seconds per 1000 trajectory recoveries, whereas the competitor with runner-up quality costs 18.17 seconds; \algo trains in 5.49 minutes per epoch, while the competitor needs 109.7 minutes per epoch.

In summary, we make the following  contributions. 
\begin{itemize}[leftmargin=*]
    \item We methodically study two related research problems, map matching and trajectory recovery on road networks, and develop efficient and accurate methods \gts and \algo. 
    \item In \gts for map matching, we formulate the task of mapping a GPS point to a segment as a classification problem over a small set of candidate segments, and develop  novel embedding techniques to align sparse trajectories to routes.
    \item For sparse trajectory recovery, \algo focuses on the segments in the routes returned by \gts and captures the intrinsic patterns of trajectories and routes via effective encoding and decoding processes to predict missing points.
    \item The superiority of \algo and \gts, in terms of efficiency and effectiveness, is evaluated over massive  trajectory data on road networks.  
\end{itemize}

\section{Problem Formulation}\label{sec:preliminary}

\begin{definition}[Road Network]
A road network is modeled as a directed graph $G=(V,E)$, where $V$ is the set of nodes, and $E$ is the set of directed edges.
A node $v\in V$ represents an intersection or a road end.  A directed edge $e=(u,v)\in E$ is a road segment from entrance node  $u$ to exit node $v$.
Let $\numseg=|E|$ and $\numinter=|V|$ denote the number of road segments and intersections, respectively. 
\end{definition}

\begin{definition}[Trajectory] 
A trajectory $\tgps$ is a sequence of GPS points with timestamps, \ie $\tgps=\langle \pgps_1,\pgps_2,...,\pgps_\lt\rangle$, where  $\lt$ is the sequence length. A GPS point is $\pii=\langle\textit{lat, lng},t_i\rangle$ consisting of latitude $lat$, longitude $lng$ and timestamp $t_i$.  
\end{definition}

\fig~\ref{fig:eg-rnet} shows a trajectory $\tgps=\langle \pgps_1,\pgps_2,\pgps_3\rangle$ with three GPS points (black dots) and a road network $\rnet$.
Note that a GPS point $\pii$ may not lie exactly on any road segment in  $\rnet$.

\begin{definition}[Route]\label{def:route}
A route $\Route$ on road network $\rnet$ is a sequence of road segments forming a \textit{path} on $\rnet$. Typically, the consecutive road segments in $\Route$ are different. 
\end{definition}

\fig~\ref{fig:eg-rnet} shows a route $\Route=\langle e_1,e_2,e_4,e_6\rangle$ (blue arrows).
The problem of map matching is defined below. 

\begin{definition}[\textbf{Map Matching}\cite{GraphMM24TKDE}]\label{def:mapmatching}
Given a road network $\rnet$ and a trajectory $\tgps$, map matching is to map the GPS points in $\tgps$ onto the road segments of $\rnet$ to find the route $\Route$ of $\tgps$.
\end{definition}

In \fig~\ref{fig:eg-rnet}, route $\Route=\langle e_1,e_2,e_4,e_6\rangle$ is the map-matched result of trajectory $\tgps$, where $\pgps_1,\pgps_2,\pgps_3$ are mapped to segments $e_1,e_4, e_6$ respectively, with $e_2$  connecting $e_1$ and $e_4$.

In what follows, we explain  the definitions related to trajectory recovery on road network.
A point on a road segment, referred to as a map-matched point, is defined as follows. 
\begin{definition}[Map-matched point \cite{MTrajRec21}]\label{def:mapmatchedpoint}
A map-matched point $\pjs=(e,r,t)$ denotes a point on segment $e$ at time $t$, with position ratio $r \in [0,1)$ representing $\pjs$'s relative position from the entrance of segment $\pjs.e$ to its total length. 
\end{definition}

\fig~\ref{fig:eg-rnet} shows 5 map-matched points  (red crosses), \eg  $\pseg_3=\langle e_4, 0.4, t\rangle$ on segment $e_4$ with position ratio $r=0.4$. 
Given the GPS coordinates of the entrance and exit of a segment $\pjs.e$, it is easy to get the GPS of $\pjs$ by interpolation. 

{A \textit{sampling rate} $\eps$ is the time interval between two consecutive points of a trajectory.
A \textit{sparse trajectory}  has a low sampling rate, \ie long intervals between consecutive points. For example, trajectory $\tgps=\langle\pgps_1,\pgps_2,\pgps_3\rangle$ shown as black dots in \fig~\ref{fig:eg-rnet} illustrates a sparse trajectory.}

\begin{definition} [Map-Matched $\eps$-Sampling Trajectory $\trajhigh$~\cite{RNTrajRec23,MTrajRec21}]
A map-matched trajectory with $\eps$-sampling rate, $\trajhigh$, is a sequence of map-matched points in Definition~\ref{def:mapmatchedpoint}, \ie 
$\trajhigh=\langle \pseg_1,\pseg_2,...,\pseg_{\lthigh}\rangle$, where $\pjs=(e,r,t)$ and time interval between consecutive points satisfies $\pseg_{j+1}.t-\pseg_{j}.t=\eps, \text{for }1\leq j<{\lthigh}$.  

\end{definition}

In \fig~\ref{fig:eg-rnet},  $\thigh=\langle\pseg_1,\pseg_2,\pseg_3,\pseg_4,\pseg_5\rangle$ serves as an example of map-matched $\epsilon$-sampling trajectory.
Then, trajectory recovery on road network is defined below.

\begin{definition}[\textbf{Trajectory Recovery on Road Network}~\cite{RNTrajRec23,MTrajRec21}]
Given a sparse GPS trajectory  $\tgps=\langle \pgps_1,\pgps_2,...\pgps_\lt\rangle$, and a target high sampling rate $\eps$, trajectory recovery aims to recover the map-matched $\eps$-sampling trajectory $\thigh=\langle \pseg_1,\pseg_2,...,\pseg_{\lthigh}\rangle$. 
\end{definition}

For instance, in \fig~\ref{fig:eg-rnet}, for  the sparse trajectory $\tgps=\langle p_1,p_2,p_3\rangle$, the goal is to recover its map-matched $\eps$-sampling trajectory $\thigh=\langle \pseg_1,\pseg_2,\pseg_3,\pseg_4,\pseg_5\rangle$.

\header 
{\textbf{Differences between the Two Tasks.}
Trajectory recovery involves  inferring  missing points of a sparse trajectory to meet the target $\eps$ sampling rate and  matching all points onto the road network $\rnet$ simultaneously.
In \fig~\ref{fig:eg-rnet}, $\pseg_2$ and $\pseg_3$ in $\thigh$ are the inferred points to satisfy the $\eps$ sampling rate, while $\pseg_1,\pseg_4, \pseg_5$ are map-matched for  GPS points $\pgps_1,\pgps_2,\pgps_3$ of the sparse trajectory $\tgps$.
In contrast, map matching aims to find the route $\Route=\langle e_1,e_2,e_4,e_6\rangle$ of $\tgps$.
Another difference is that the route $\Route$ from map matching is a  path on $\rnet$ from a source to a destination, whereas segments of map-matched points in $\thigh$ can be disconnected, with consecutive points possibly on the same segment. For instance, in \fig~\ref{fig:eg-rnet}, $\thigh=\langle\pseg_1,\pseg_2,\pseg_3,\pseg_4,\pseg_5\rangle$ has segment sequence $\langle e_1,e_4,e_4,e_4,e_6\rangle$, where $e_1$ and $e_4$ are not connected and $\pseg_2,\pseg_3,\pseg_4$ are all on segment $e_4$.}
\tab~\ref{tbl:notations} shows the frequently used notations in the paper.

\begin{table}[!t]
\centering
\renewcommand{\arraystretch}{1.1}
\begin{small}
\caption{Frequently used notations}\vspace{-1mm} \label{tbl:notations}
\resizebox{1.01\columnwidth}{!}{%
	\begin{tabular}{|p{1.00in}|p{2.30in}|}
 
		\hline
		{\bf Notation} &  {\bf Description}\\
  		\hline
		$\trajdata$ & Historical trajectory data $\trajdata$.\\
  \hline
		$\rnet=(V,E)$, $\numseg$ & A road network $\rnet$ with road segment set $E$ and intersection set $V$. $n$ is the number of segments $|E|$.\\
		\hline
		$e=(u,v)$ & A road segment $e$ from node $u$ to node $v$. \\
            \hline 
            $\pii$ & A GPS point with coordinate and timestamp\\ \hline 
            $\pis=(e,r,t)$ & A map-matched point on  segment $e$ with position ratio $r$ and time $t$\\ \hline
            $\tgps=\langle p_1,...,p_\lt\rangle$ & A sparse trajectory with $\lt$ GPS points.\\
            \hline 
            $\Route,\lR$ & A route, \ie a path with $\lR$ segments. \\
  \hline
  $\eps$ & The target sampling rate\\
		\hline $\thigh=\langle\pseg_1,...,\pseg_{\lthigh}\rangle$ & The recovered map-matched $\eps$-sampling trajectory with sequence length $\lthigh$.\\
		\hline
		
		$\candset_{p_i}$, $\candsize$ & The candidate segment set of GPS point $\pii$ with size $\candsize$.\\
		\hline
  $P(\cj|\pii)$ & The probability that $\cj$ is the segment of $\pii$\\
		\hline
    $\mathbf{M}, \mathbf{M}[i,:], \mathbf{v}$ & A matrix $\mathbf{M}$, its $i$-th row vector $\mathbf{M}[i,:]$, and a vector $\mathbf{v}$. \\ \hline 
	\end{tabular}%
}
\end{small}
\vspace{-3mm}
\end{table}

\section{Related Work}\label{sec:relatedwork} 
\noindent
\textbf{Trajectory Recovery on Road Network.}
While trajectory recovery on road networks is related to map matching, recent studies often develop specialized end-to-end methods by learning from historical trajectories~\cite{MTrajRec21,RNTrajRec23,STGED24}.
MTrajRec~\cite{MTrajRec21} introduces a sequence-to-sequence model with multitask learning to interpolate missing points in sparse trajectories while adhering to road network constraints. RNTrajRec~\cite{RNTrajRec23} considers road network structure and the surrounding subgraphs of observed GPS points for trajectory recovery. MM-STGED~\cite{STGED24} presents a graph-based encoder-decoder for trajectory recovery. 
The  methods in~\cite{RNTrajRec23,MTrajRec21} typically project the predicted locations of missing points across all segments in a road network, which can be large. 
Instead, our method \algo only recovers the missing points on the segments in the route returned by our map-matching method \gts. 
The segments in the route are much fewer than the segments in $\rnet$, rendering the efficiency of our designs, as demonstrated in the experiments.

\header
\textbf{Trajectory Recovery in Free Space.} 
Another set of studies focuses on recovering trajectories in free space~\cite{DHTR21, MuslehM23,ElshrifIM22, TERI23, AttnMove21, PeriodicMove21, XiZLGXH19}, such as social media check-in data, without the constraints of road networks. Elshrif~\etalstd~\cite{ElshrifIM22} introduce a heuristic search algorithm to estimate locations between two distant consecutive points within a trajectory. DHTR~\cite{DHTR21} utilizes a BiLSTM with spatial and temporal attention mechanisms and incorporates a Kalman filter-based calibration component to reconstruct missing GPS points.  Chen~\etalstd~\cite{TERI23} develop TERI, a method for trajectory recovery with irregular time intervals.
Xia~\etalstd~\cite{AttnMove21} employ  self-attention mechanisms within and across trajectories to infer unobserved locations. Sun~\etalstd~\cite{PeriodicMove21} integrate a graph neural networks to model individual trajectories to capture transition patterns. These studies focus on free space, and exhibit suboptimal performance when adapted to road network constraints.

\header 
\textbf{Map Matching.} There is a rich collection of studies on map matching~\cite{PinkH08,HMM09,STMM09,fmm2018,HRIS12,NMM14,AntMapper18, DeepMM22TMC,DMM24TKDE,GraphMM24TKDE, LHMM24ICDE, L2MM23TKDD}.
A pioneer study~\cite{HMM09} leverages HMM to find the most likely road route. Zheng~\etalstd~\cite{HRIS12} introduce a history-based route inference system. 
Lou~\etalstd~\cite{STMM09} construct a candidate graph using the spatial and topological structures of the road network and the temporal features of trajectories. 
These methods usually exhibit degraded performance with sparse GPS trajectory data~\cite{HMM09}. 
FMM~\cite{fmm2018} enhances HMM with a set of acceleration mechanisms. 
A recent trend is to extract patterns from historical trajectories to conduct map matching~\cite{DeepMM22TMC,DMM24TKDE,GraphMM24TKDE, LHMM24ICDE, L2MM23TKDD}.
LHMM~\cite{LHMM24ICDE} enhances the HMM model by incorporating knowledge obtained from neural networks into learned probabilities, while DMM~\cite{DMM24TKDE} adopts a recurrent neural network with a reinforcement learning scheme.
GraphMM~\cite{GraphMM24TKDE} is a graph-centric approach to incorporate graph neural networks and conditional models to leverage road and trajectory graph topology for map matching. 
DeepMM~\cite{DeepMM22TMC} is an end-to-end deep learning method with statistical and graph-based data augmentation techniques. 
{However, these methods incur significant computational costs for training and inference, as shown in experiments. Overall, existing map matching methods struggle with effectiveness, efficiency, or both when dealing with sparse trajectories.}

\header \textbf{Trajectory Representation Learning.}
A plethora of studies have focused on embedding trajectory data into low-dimensional vectors~\cite{tvec18, NeuTraj19, T3S21, TrajGAT22, TrajCL23, STVec22, GTS21}. These embeddings encapsulate both the spatial and sequential characteristics of trajectories and can serve as input of trajectory recovery tasks. 
TrajGAT~\cite{TrajGAT22} introduces a graph-based attention to capture long-term dependencies by representing each trajectory as a graph. TrajCL~\cite{TrajCL23} proposes a dual-feature, self-attention-based trajectory encoder that adaptively incorporates both structural and spatial features. 
ST2Vec~\cite{STVec22} produces time-aware trajectory representations by integrating spatial and temporal elements.
In our experiments, we adopt these methods~\cite{TrajGAT22, TrajCL23, STVec22} with a decoding process for trajectory recovery. However, these approaches often yield suboptimal quality in the experiments, highlighting the need for specialized technical designs tailored to trajectory recovery on road networks.

\section{\textsf{MMA}:  Map Matching}\label{sec:gps2seg}

As mentioned, our trajectory recovery method \algo utilizes the route $\Route$ of trajectory $\tgps$ obtained by our map matching method \gts to recover the map-matched $\eps$-sampling trajectory $\thigh$ of $\tgps$.
Hence, we first present \gts in this section, while developing \algo in Section~\ref{sec:rec}.

Given a sparse trajectory $\tgps=\langle \pgps_1,...,\pgps_\lt\rangle$, the major task of \gts is mapping every GPS point $\pii$ to its corresponding segment  $e_i$ in road network $\rnet$, such as mapping $\pgps_2$ to $e_4$ in \fig~\ref{fig:eg-rnet}. 
Due to the sparsity of $\tgps$, this task is non-trivial.
After conducting an empirical analysis, we formulate it as a classification problem over a small candidate set of segments near $\pii$, rather than expensively considering all segments in $\rnet$ (Section~\ref{sec:gtsconcepts}). Then we develop point embedding and candidate segment embedding techniques in \gts to exploit the sequential nature of $\tgps$, the properties of candidate segments, and the characteristics of road network for effective classification (Section~\ref{sec:gtsArchi}). Once the GPS points of $\tgps$ are accurately mapped  onto segments, \gts adopts elementary route planning methods to link disconnected segments into a route $\Route$ to return.

\subsection{\gts Formulation}\label{sec:gtsconcepts}

Intuitively, the road segment where $\pii$ is located should be in its vicinity, making it unnecessary to consider all road segments in $\rnet$, which can be numerous. However, the nearest segment to $\pii$ may not always be the one where $\pii$ actually resides. We conduct an empirical analysis to validate.

\begin{figure}[!t]
\centering
\begin{small}
\begin{tikzpicture}
    \begin{customlegend}[legend columns=12,
	    legend entries={PT,XA,BJ,CD},
	    legend style={at={(0.45,1.15)},anchor=north,draw=none,font=\scriptsize,column sep=0.05cm}]
        
	    \addlegendimage{line width=0.25mm,color=violet}
	    \addlegendimage{line width=0.25mm,color=cyan}
	    \addlegendimage{line width=0.25mm,color=orange}
	    \addlegendimage{line width=0.25mm,color=blue} 
    \end{customlegend}
\end{tikzpicture}
\\[-\lineskip]
\vspace{-2mm}

		\hspace{-6mm} \begin{tikzpicture}[scale=1]
		\begin{axis}[
		height=\columnwidth/2.5,
		width=\columnwidth/1.6,
		ylabel={\footnotesize\em ratio},
		xmin=1, xmax=10,
		xtick={1,2,3,4,5,6,7,8,9,10},
		every axis y label/.style={at={(current axis.north west)},right=3mm,above=0mm},
		every axis x label/.style={at={(current axis.south)},above=-7mm},
		label style={font=\small},
		tick label style={font=\scriptsize},
		]
		
		\addplot[line width=0.25mm,color=violet]
		table[x=k, y=pt, col sep=comma] {data/top0.tex};
		
		\addplot[line width=0.25mm,color=cyan]
		table[x=k, y=xa, col sep=comma] {data/top0.tex};
		
		\addplot[line width=0.25mm,color=orange]
		table[x=k, y=bj, col sep=comma] {data/top0.tex};
		
		\addplot[line width=0.25mm,color=blue]
		table[x=k, y=cd, col sep=comma] {data/top0.tex};
		\end{axis}
		\end{tikzpicture} \hspace{-2mm}%

\vspace{-1mm}
\end{small}
\caption{The ratio of GPS points with their segments in their top-$\candsize$ nearest segments when varying $\candsize$ from 1 to 10.} 
\label{fig:exp-knn}
\vspace{-2mm}
\end{figure}
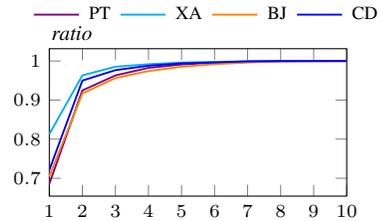

Given historical trajectory data $\trajdata$ for training, for every GPS point $\pii$ in every trajectory  in $\trajdata$, we obtain its top-$\candsize$ nearest segments in road network $\rnet$, and calculate the ratio of all points with their ground-truth road segment  among their top-$\candsize$ nearest segments. 
\fig~\ref{fig:exp-knn} reports the ratio curves when varying $\candsize$ from 1 to 10 across the four datasets used in our experiments (data description in Section~\ref{subsec:setup}).
When $\candsize=1$, the ratios of GPS points with their nearest segments as their actual segments are relatively low on all datasets, approximately 0.7 for three of the four datasets.
This indicates the inaccuracy of solely relying on the nearest segment of a GPS point for map matching. 
More importantly, observe that when $\candsize$ increases to 10, the ratio values approach 1 across all datasets. 
This supports the intuition that the segment of a GPS point is typically nearby. Specifically, with probability close to 1, the true segment of a GPS point is within the top-$\candsize$ nearest segment (\eg $\candsize=10$).

{According to~\cite{renfro2021analysis}, civilian GPS devices have horizontal errors of 7.0\textit{m} at a 95\% confidence level and 30\textit{m} at a 99\% confidence level. For the datasets in Fig. 2, the average distance between a GPS point and its 10th nearest segment is 82.36\textit{m} (PT), 122.03\textit{m} (XA), 90.81\textit{m} (BJ), and 95.85\textit{m} (CD). These exceed the 30\textit{m} threshold, confirming that the actual segment is within the top-$\candsize=10$ nearest segments with high probability.
We also calculated the average distance to the true segment (and nearest segment): 5.56(4.67)\textit{m}, 4.96(4.23)\textit{m}, 6.84(5.21)\textit{m}, and 4.20(3.16)\textit{m} for PT, XA, BJ, and CD, respectively. These distances are within the 7.0\textit{m} range at the 95\% confidence level, consistent with the observation  that when 
$\candsize=1$, the ratio of GPS points with their nearest segments as actual segments is about 0.7 for 3 out of 4 datasets.}

Therefore, we only need to investigate these segments as candidates  to identify the segment of $\pii$, without tediously looking at all segments in $\rnet$, where $\candsize \ll |V|$. In Definition~\ref{def:gps-mask}, we define the candidate segment set $\Cpi$ of $\pii$ as the set of its top-$\candsize$ nearest segments in $\rnet$. 

\begin{definition}[Candidate Segment Set of a GPS Point] \label{def:gps-mask} 
Given a road network $\rnet$ and a  GPS point $\pii$, the candidate segment set $\Cpi$ contains the top-$\candsize$ nearest segments ranked by the perpendicular distance from $\pii$ to the segments in  $\rnet$. 
\end{definition}

The usage of the candidate set $\Cpi$ of $\pii$ is two-fold. Besides acting as the candidate pool, the semantics of all segments in  $\Cpi$  also server  as the context of $\pii$ to enhance its embedding, to be explained shortly. We can efficiently get  $\Cpi$ of $\pii$  via a top-$\candsize$ query of $\pii$ over an R-tree~\cite{rtree97} index of road segments.

After defining the candidate segment set $\Cpi$ for a GPS point $\pii$, we propose to formulate the task of finding $\pii$'s segment as a classification problem over the candidates in $\Cpi$.
Below, for a GPS point $\pii$, we state the class labels of the candidate segments in $\Cpi$ and the classification problem. 

\header 
\textbf{Class Labels in $\Cpi$.} For a GPS point $\pii$  of a trajectory, a candidate segment $c\in\Cpi$ has class label $y_{c}=1$, if $c$ is the ground-truth segment of $\pii$; otherwise, the class label of $c$ is $y_c=0$.
At most one segment in $\Cpi$ has class label 1. It is possible that all candidates are in class 0 if the ground truth is not included in $\Cpi$, which is rare when $\candsize=10$.

\header \textbf{Problem Statement.}
Given training trajectory data $\trajdata$ on road network $\rnet$, the goal is to train  a classification model to accurately identify the true segment $e_i$ (class label 1) of point $\pii$ from its candidate set $\Cpi$. 
\gts  computes the probability $\keyprob(c|\pii)$ for every candidate segment $c\in\Cpi$ and the candidate with the highest probability is the segment of $\pii$.

\subsection{\gts Method}\label{sec:gtsArchi}

\begin{figure}[t]
	\centering	\includegraphics[width=1\columnwidth]{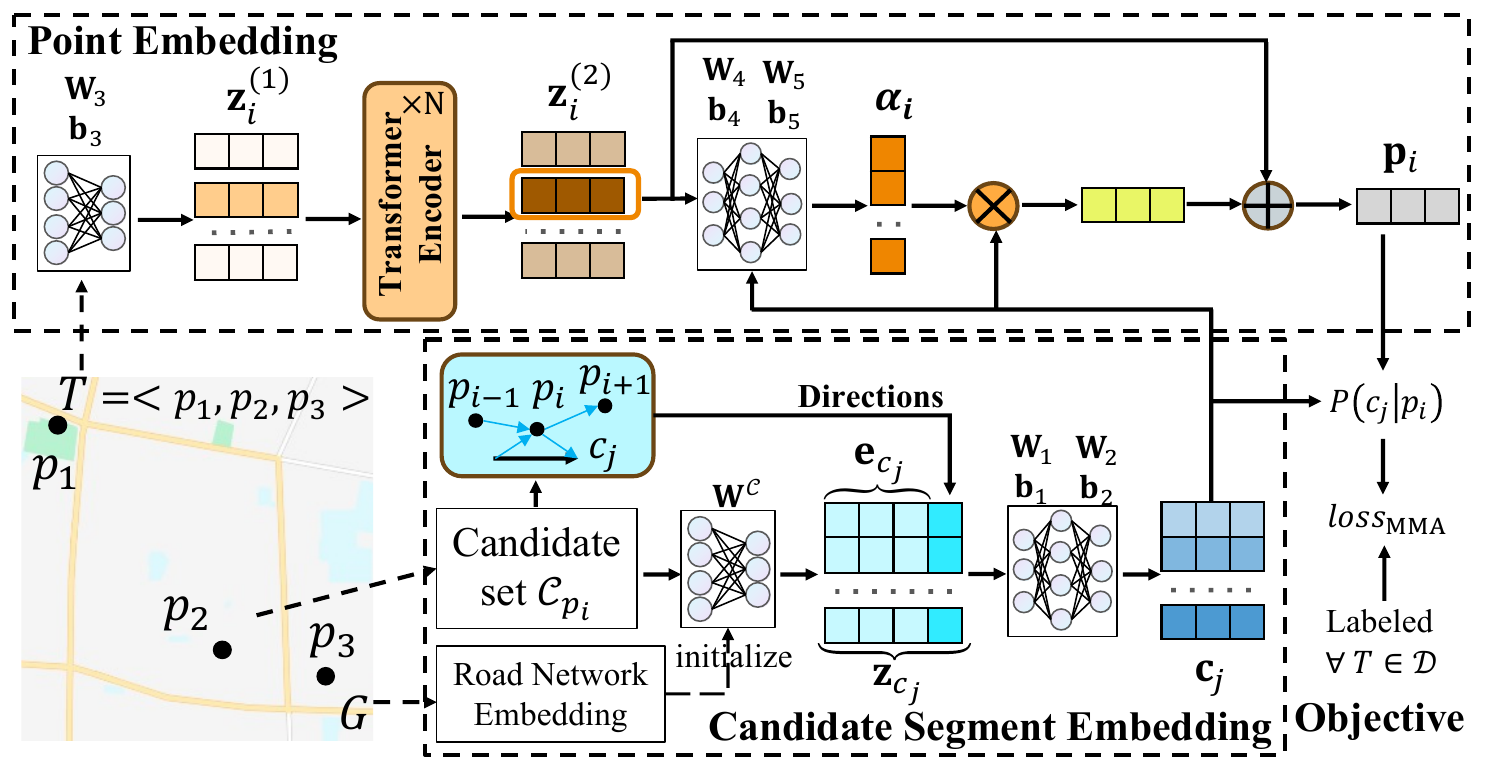}
	\vspace{-5mm} \caption{Map GPS Points to Segments}
	\label{fig:archi}
	\vspace{-3mm}
\end{figure}

In Algorithm~\ref{alg:gts}, \gts first maps every GPS point $\pii$ in $\tgps$ to a segment $e_i$ (Lines 1-9) and then constructs route $\Route$ (Lines 10-13). 
As mentioned, we formulate the task of mapping $\pii$ to a segment as a classification problem over $\Cpi$. To achieve accurate classification,  we design a point embedding module and a candidate segment embedding module, and adopt a cross entropy loss as objective, as illustrated in \fig~\ref{fig:archi}.

\header \textbf{Candidate Segment Embedding.} 
As shown in the bottom part of \fig~\ref{fig:archi}, we aim to generate an effective embedding $\cjemb$ for each candidate segment $\cj\in\Cpi$ of point $\pii$ in the sparse trajectory $\tgps$. The embedding $\mathbf{c}_j$ should intuitively reflect the connectivity of $\cj$ in the road network $\rnet$  and  its relationship with point $\pii$. To preserve the first aspect, we use network embedding Node2Vec~\cite{node2vec16} to pre-learn segment embeddings $\nvmatorigin\in \mathbb{R}^{\numseg \times d_0}$ for all $n$ segments in $\rnet$. 
Then we initialize the learnable parameters $\mathbf{W}^\candset$ of a fully-connected (FC) layer with these embeddings. 
This FC layer is subsequently trained to produce the representations $\mathbf{e}_{\cj}$ of $\cj$ in $\rnet$. 
For every $\cj\in\Cpi$, the FC layer transforms its one-hot id vector $\onehot_{\cj}$ to a dense representation $\mathbf{e}_{\cj}$ via learnable weights $\mathbf{W}^\candset$,
\begin{equation}\label{eq:fccand}
\mathbf{e}_{\cj}=\onehot_{\cj}\mathbf{W}^\candset
\end{equation}
where $\mathbf{W}^\candset \in \mathbb{R}^{\numseg \times d_0}$ contains the learnable parameters and initially $\mathbf{W}^\candset=\nvmatorigin$, and $\numseg$-dimensional one-hot vector $\onehot_{\cj} \in \{0,1\}^{\numseg}$ has all elements are 0, except 1 at the position corresponding to the id of $\cj$.

\begin{algorithm}[!t]
	\caption{\gts for Map Matching}
	\label{alg:gts}
 \small
	\KwIn{A sparse trajectory 
 $\tgps=\langle p_1,...,p_\lt\rangle$, road network $\rnet$}  
	\KwOut{Route $\Route$}
        Get embedding $\zic$ for every $\pii$ in Eq.\eqref{eq:te-point} preserving sequential patterns in $\tgps$\;
        \ForEach{GPS point $\pii$ in $\tgps$}{
            Get candidate set $\Cpi$ with size $\candsize$ for $\pii$ by Definition~\ref{def:gps-mask}\;
            \ForEach{candidate segment $\cj\in\candset_{\pii}$}{
                $\cjemb\gets$ Candidate Segment Embedding by Eq.\eqref{eq:fccand}\eqref{eq:mlpcand}\;
                $\alpha_{j,i}\gets$ Eq.\eqref{eq:attn} with $\zic$ and $\cjemb$\;
            }
            $\piemb \gets \zic+ \sum_{\forall \cj\in\Cpi} \alpha_{j,i}\cdot\cjemb$\;
            $\forall \cj\in\Cpi,\keyprob(\cj|\pii)\gets sigmoid(\cjemb \cdot \piemb)$\;
            {$\pii$ is mapped to segment $e_i\gets \arg\max_{\cj\in\candset_{\pii}}P(\cj|\pii)$\;}
        }
        
        $\Route\gets\langle\rangle$\;
        \For{$i=1,...,\lt-1$}{
         Get the route $\Route_i$ between $\pii$'s segment $e_i$ and $\pgps_{i+1}$'s segment $e_{i+1}$\;
         $\Route$.append$(\Route_i)$\;
         }
    \Return $\Route$\;
\end{algorithm}

For the second aspect, we consider the directional relationship between the candidate $\cj\in\Cpi$ and $\pii$ in the sparse trajectory $\tgps$.
Specifically, as shown in \fig~\ref{fig:archi}, for every segment $\cj$, treated as a vector from its entrance to its exit, we compute four cosine similarity values: its similarity with the vector from the entrance of $\cj$ to $\pii$, from $\pii$ to the exit of $\cj$,  from $p_{i-1}$ to $\pii$ in $\tgps$, and  $\pii$ to $p_{i+1}$ in $\tgps$.
We concatenate these four  similarity values and  with $\mathbf{e}_{\cj}$ to get the embedding $\mathbf{z}_{\cj}\in \mathbb{R}^{(d_0+4)}$.
This embedding $\mathbf{z}_{\cj}$ is then fed into a Multi-layer Perceptron (MLP) in  Eq.\eqref{eq:mlpcand}  to produce the final candidate embedding $\mathbf{c}_j \in \mathbb{R}^{d_2}$. 
Briefly, MLP is a fully connected feed-forward neural network, and it often combines with nonlinear activation function, \eg $ReLU(x)=\max(0,x)$, to bring non-linearity into the model, to alleviate the vanishing gradient problem~\cite{relu11}.
\begin{equation}\label{eq:mlpcand}
    \mathbf{c}_j = ReLU(\mathbf{z}_{\cj} \mathbf{W}_1 +\mathbf{b}_1) \mathbf{W}_2 +\mathbf{b}_2
\end{equation}
where $\mathbf{W}_1\in \mathbb{R}^{(d_0+4)\times d_1}$, $\mathbf{b}_1\in\mathbb{R}^{d_1}$, $\mathbf{W}_2\in\mathbb{R}^{d_1\times d_2}$, and $\mathbf{b}_2\in\mathbb{R}^{d_2}$.

As shown in \fig~\ref{fig:archi}, the candidate embedding  $\cjemb$ serves  two purposes:  training the classification objective, and used as the context of $\pii$ to enhance point embedding $\piemb$ developed below. 

\header
\textbf{Point Embedding.}  
In the top part of \fig~\ref{fig:archi}, we generate  the  embedding $\piemb$ of a GPS point $\pii$ by considering the GPS information of $\pii$, the sequential features of $\tgps$, and the embeddings of its candidate segments in $\Cpi$. The idea is that each  GPS point $\pii$ has a unique set of candidate segments $\Cpi$, which should be reflected in the embedding $\piemb$.
Initially, point $p_i$ has a vector $\zia$ containing its min-max normalized latitude, longitude and timestamp.
$\zia$ is fed into an FC to get $\zib= \zia \mathbf{W}_3+\mathbf{b}_3$, where $\mathbf{W}_3\in \mathbb{R}^{3\times d_2}$, $\mathbf{b}_3\in\mathbb{R}^{d_2}$ are the learnable parameters.

Let $\Zb=[\zb{1}, \zb{2}, ..., \zb{\lt}]$ be the embeddings of all GPS points in $\tgps$. As depicted in \fig~\ref{fig:archi}, we adopt a transformer~\cite{trans17} to convert $\Zb$ to $\Zc=[\zc{1}, \zc{2}, ..., \zc{\lt}]$, capturing the sequential features in $\tgps$ with two stacked transformer layers,  
\begin{equation} \label{eq:te-point}
\Zc =\text{Trans}(\Zb)
\end{equation}

We explain the details of a transformer~\cite{trans17}, which will also be used in Section~\ref{sec:rec}. 
A transformer layer comprises two sub-layers: a multi-head self-attention mechanism and a position-wise feed-forward network (FFN). Self-attention allows the model to focus on relevant parts of the input sequence selectively. Multi-head attention (MHAttn) in Eq.\eqref{eq:mhattn} maps the input sequence representations to output representations using a scaled dot-product function across multiple heads:
\begin{equation} \label{eq:mhattn}
    {\small
    \begin{aligned}
        &\text{MHAttn}(\mathbf{Q}, \mathbf{K}, \mathbf{V}) = \text{concat}[head_1, ..., head_h] \cdot \mathbf{W}^O \\
        &head_i  = \text{Attention}(\mathbf{Q}\mathbf{W}_i^Q,\mathbf{K}\mathbf{W}_i^K,\mathbf{V}\mathbf{W}_i^V) \\
        &\text{Attention}(\mathbf{Q}, \mathbf{K}, \mathbf{V})  = \text{softmax}({\mathbf{Q} \mathbf{K}^T}/{\sqrt{d}}) \mathbf{V} \\
    \end{aligned}
    }
\end{equation}
where $\mathbf{Q}, \mathbf{K}$ and $\mathbf{V}$ represent the query, key, and value matrix respectively,
$\mathbf{W}_i^Q, \mathbf{W}_i^K, \mathbf{W}_i^V \in \mathbb{R}^{d_{in} \times d_{in}/h}$ are projection matrices for the $i$-th head, $h$ is the number of attention heads, and $\mathbf{W}^O\in \mathbb{R}^{d_{in}\times d_{out}}$ are parameters for the output, and $d$ is the feature dimension of $\mathbf{K}$.

An FFN is two-layer MLP with ReLU activation in Eq.\eqref{eq:ffn}.
\begin{equation} \label{eq:ffn}
    \small
    \text{FFN}(\mathbf{X}) = ReLU(\mathbf{X} \mathbf{W}_x + \mathbf{b}_x) \mathbf{W}_y  +\mathbf{b}_y
\end{equation}
where $\mathbf{W}_x$, $\mathbf{b}_x$, $\mathbf{W}_y$, $\mathbf{b_y}$ are learnable parameters.

Given input $\mathbf{X}$, a transformer layer applies MHAttn to $\mathbf{X}$ itself, followed by an FFN, both with residual connection and layer normalization in Eq.\eqref{eq:trans}.  
\begin{equation} \label{eq:trans}
{\small
\begin{aligned}
    &\text{Trans}(\mathbf{X}) = \text{LayerNorm}(\mathbf{X}\sprime + \text{FFN}(\mathbf{X}\sprime)) \\
    &\mathbf{X}\sprime  =\text{LayerNorm}(\mathbf{X} + \text{MHAttn}(\mathbf{X}, \mathbf{X}, \mathbf{X})) 
\end{aligned}}
\end{equation}
where MHAttn is defined in Eq.\eqref{eq:mhattn} and FFN is Eq.\eqref{eq:ffn}.

As explained, the candidate segments $\Cpi$ of $\pii$ represent the specific context of $\pii$, and this should be reflected in the embedding of $\pii$. 
To achieve this, we design an attention mechanism to learn the attention score $\alpha_{j,i}$ of every candidate embedding $\cjemb$ relative to $\zic$ of $\pii$. This is done in Eq.\eqref{eq:attn} via a two-layer MLP over the concatenation of $\zic$ and $\cjemb$ to get intermediate $s_{j,i}$, which is normalized by softmax normalization to get $\alpha_{j,i}$.
\begin{equation} \label{eq:attn}
    \begin{aligned}
            s_{j,i} & = ReLU(\text{concat}[\zic, \cjemb] \mathbf{W}_4 + \mathbf{b}_4) \mathbf{W}_5 + \mathbf{b}_5\\
        \alpha_{j,i} & = \frac{exp(s_{j,i})}{\sum_{k=1}^{\candsize} exp(s_{k,i})} 
    \end{aligned}
\end{equation}
where $\mathbf{W}_4\in \mathbb{R}^{2d_2 \times d_3}$, $\mathbf{b}_4\in \mathbb{R}^{d_3}$, $\mathbf{W}_5\in \mathbb{R}^{d_3 \times 1}$, $\mathbf{b}_5\in \mathbb{R}^{1}$ are the learnable parameters.

The attention scores  are  then used to aggregate all $\cjemb$ of $\cj\in\Cpi$ and $\zic$, to get the final embedding $\piemb$ of point $\pii$,  
\begin{equation} \label{eq:pointembedding}
        \textstyle \piemb = \zic+ \sum_{\forall \cj\in\Cpi} \alpha_{j,i}\cdot\cjemb
\end{equation}

Given the embedding $\piemb$ of point $\pii$ and the embedding $\cjemb$ of a candidate segment $\cj\in\Cpi$, the probability $\keyprob(\cj|\pii)$ that $\cj$ is the segment of $\pii$ is computed as the inner product of $\cjemb$ and $\piemb$  normalized by $sigmoid(x)=\frac{1}{1+exp (-x)}$ into $(0,1)$.
\begin{equation}\label{eq:probmm}
\keyprob(\cj|\pii)=sigmoid(\cjemb \cdot \piemb)
\end{equation}

\header 
\textbf{Objective.} During offline training stage,  a candidate $\cj \in\Cpi$ has class label $y_{\cj}=1$ or $0$.
\gts employs a binary cross entropy loss for a point $\pii$, 
\begin{equation*}
    \small
    L_{\pii}= -\sum_{\forall \cj \in \candset_{p_i}} \left(y_{\cj} \log \keyprob(\cj|p_i)+(1-y_{\cj}) \log \left(1-\keyprob(\cj|p_i)\right)\right)
\end{equation*}

On training data $\trajdata$, the overall training loss of \gts is, 
\begin{equation}
        \label{eq:loss}
\textstyle L_{\gts}=\frac{1}{|\trajdata|} \sum_{\forall \tgps \in \trajdata} \sum_{\forall p_i \in \tgps} L_{\pii}
\end{equation}

\header
\textbf{Algorithm.}
Algorithm~\ref{alg:gts} shows the pseudocode \gts for forward execution to recover the route $\Route$ of a sparse trajectory $\tgps$ on road network $\rnet$.
From Lines 1-9, \gts maps each GPS point $\pii$ to its corresponding segment.
At Line 1, we first incorporate the sequential pattern in $\tgps$ into embedding $\zic$ for every GPS point $\pii$, to be used later.
Then, for every $\pii$, we obtain its candidate segment set $\Cpi$ and generate candidate segment embedding $\cjemb$ for each $\cj\in\Cpi$ (Lines 2-5).
The attention between $\pii$ and each candidate segment $\cj$ is obtained via Eq.\eqref{eq:attn} with $\zic$ and $\cjemb$ as input (Lines 6).
Using the attention scores, we derive the final point embedding $\piemb$ via Eq.\eqref{eq:pointembedding} (Line 7).
Then we compute probability $\keyprob(\cj|\pii)$ using embeddings $\cjemb$ and $\piemb$ (Line 8), and map $\pii$ to the segment with the highest probability in $\Cpi$ (Line 9).
Since the matched segments of two consecutive points $\pii$ and $\pgps_{i+1}$ may not be connected, we construct the routes $\Route_i$ between the segments of all possible $\pii$ and $\pgps_{i+1}$ when necessary (Line 10-12) and combine them to get the final $\Route$ (Line 13). 
Note that \gts is orthogonal to the route planning method used at Line 12. In our experiments, for our methods and baselines requiring route planning as a routine, we use the same DA-based method from~\cite{DRPK23} that relies on basic statistical counts.

\header 
\textbf{Complexity.} 
Let $d$ be the dimension of embeddings involved.
$\tgps$ has $\lt$ GPS points.
Line 1 costs $O(\lt^2d)$ time to run the transformer.
To get $\Cpi$ of $\pii$, a top-$\candsize$ query over R-tree takes $O(\log\numseg)$ time in average case.
MLPs at Lines 5-6 costs  $O(d^2)$ time, and $\candsize$ is constant.
Lines 2-8 are executed $\lt$ times.
The route planning routine at Line 12 has $O(l'\Tilde{deg})$ time~\cite{DRPK23}, where $l'$ is the max route length limit and $\Tilde{deg}$ is the max degree in $\rnet$, which is typically small.
Regarding $l'$ same as $\lt$, \gts costs  $O(\lt^2d+\lt d^2+\lt \log\numseg + \lt^2\Tilde{deg})$ time, where $\log\numseg$ is sub-linear to $\numseg$ and $\lt$, $d$, and $\Tilde{deg}$ are usually small, indicating the efficiency of \gts.

\section{\textsf{TRMMA}: Sparse Trajectory Recovery} \label{sec:rec}

After obtaining the route $\Route$ of a sparse trajectory $\tgps$, \algo focuses on the segments in $\Route$ to infer missing map-matched points $\pis$, thereby recovering the map-matched $\eps$-sampling trajectory $\thigh$ for $\tgps$. The segments in $\Route$ are typically much fewer than those in  $\rnet$, making \algo efficient, especially compared to existing methods~\cite{MTrajRec21,RNTrajRec23}  that evaluate all segments in $\rnet$ for trajectory recovery.

\begin{figure}[t]
	\vspace{-2mm}
	\centering	\includegraphics[width=1.02\columnwidth]{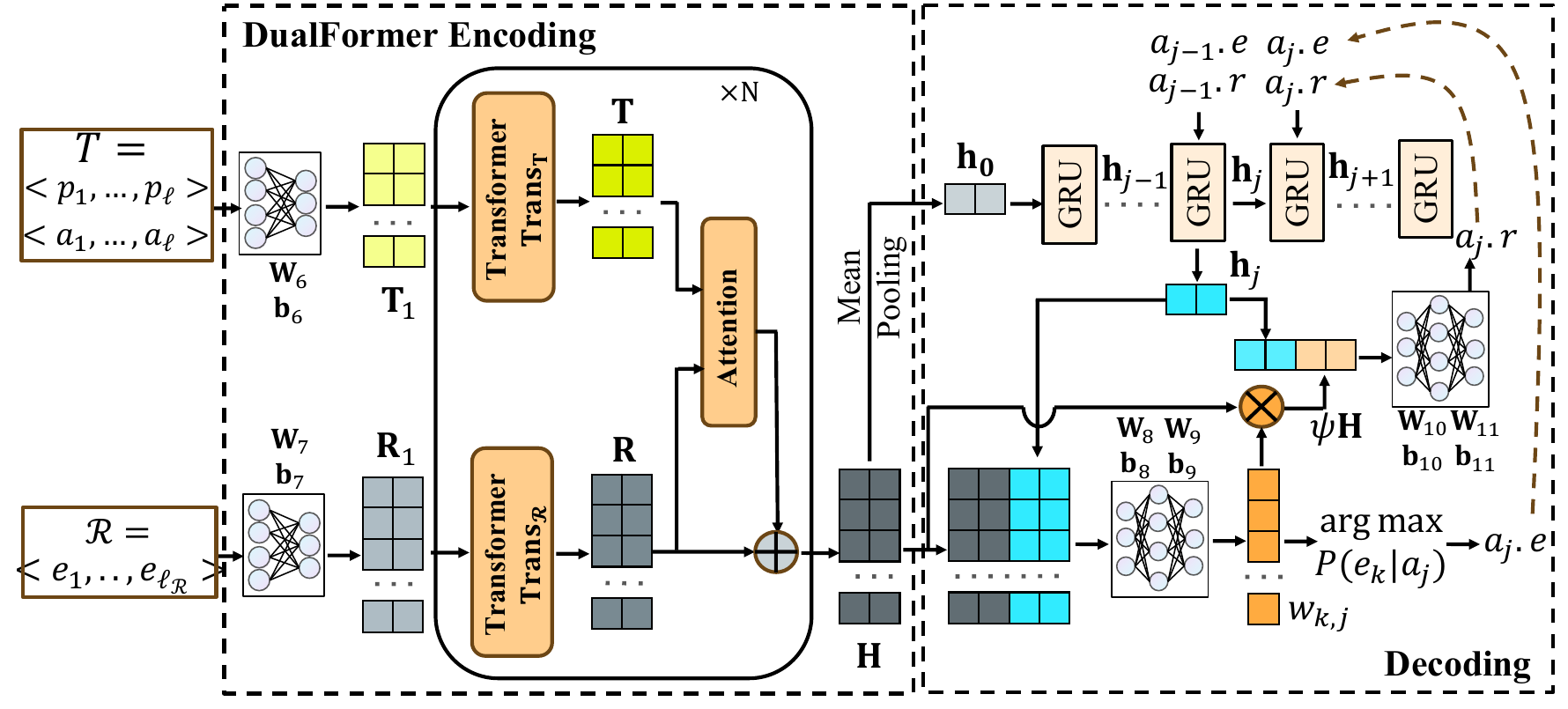}
	\vspace{-5mm} 
 \caption{\algo to Recover Missing Points on Route}
	\label{fig:archi-rec}
	\vspace{-5mm}
\end{figure}

As shown in Algorithm~\ref{alg:TR}, \algo first invokes \gts to get $\Route$ (Line 1). For each GPS point $\pii$ in $\tgps$, we get its map-matched point $\pis$ by projecting $\pii$ to its segment $e_i$ (Lines 2-4).
Then \algo utilizes the sparse trajectory $\tgps$, $\pis$ of every $\pii$ in $\tgps$, and route $\Route$ together to get $\thigh$ satisfying the target $\eps$-sampling rate from Lines 5 to 16, using a proposed DualFormer encoding module and a multitask decoding module illustrated in \fig~\ref{fig:archi-rec}.
Specifically,  we design the \enc to integrate $\tgps$ and $\Route$ sequences to produce expressive embeddings. Subsequently, we develop the \dec that utilizes the  embeddings to estimate missing points $\pjs$, including $\pjs.e$ and $\pjs.r$, to get $\trajhigh$. 
Briefly, we regard the segments in $\Route$ as the candidates to predict $\pjs.e$, while treating the estimation of $\pjs.r$ as a regression problem.

\header \textbf{\Enc.}
The \enc employs one transformer to encode the sparse trajectory $\tgps$  with sequence length $\lt$ into representations $\TM \in \mathbb{R}^{\lt \times d_h}$, and uses another transformer to encode the route $\Route$ of sequence length $\lR$ into representations $\RM \in \mathbb{R}^{\lR \times d_h}$. An attention mechanism is then applied to $\TM$ and $\RM$ to capture the intrinsic relationship between the points in $\tgps$ and the segments in $\Route$. This process yields the final embeddings $\HM \in \mathbb{R}^{\lR \times d_h}$, which serve as the input for the decoding process.

In particular, for all points $\pii$ in $\tgps$, we first initialize an embedding matrix $\Ta \in \mathbb{R}^{\lt \times d_4}$, the $i$-th row of which is the initial encoding of $\pii$ containing information of $\pii$ and its map-matched point $\pis$, including min-max normalized $\pii.lat,\pii.lng, \pii.t$, $\pis.r$ and the id embedding of segment $\pis.e$. 
$\Ta$ goes through one FC layer  to get $\Tb$, and then a transformer is applied over $\Tb$ to capture the sequential patterns of $\tgps$ into $\TM\in \mathbb{R}^{\lt \times d_h}$,
\begin{equation}\label{eq:gseq-token}   
\TM=\text{Trans}_\tgps(\Tb), \text{ } \Tb=\Ta\mathbf{W}_6+\mathbf{b}_6,
\end{equation}
where $\mathbf{W}_6\in \mathbb{R}^{d_4 \times d_h}$, $\mathbf{b}_6\in\mathbb{R}^{d_h}$ are the learnable parameters.

\begin{algorithm}[!t]

	\caption{\algo for Trajectory Recovery on Road Network}
	\label{alg:TR}
 \small
	\KwIn{ A sparse trajectory 
 $\tgps=\langle p_1,p_2,...,p_\lt\rangle$,
 a target sampling rate $\epsilon$, road network $\rnet$}  
	\KwOut{Recovered map-matched $\epsilon$-sampling trajectory $\trajhigh=\langle \pseg_1,\pseg_2,...,\pseg_{\lthigh}\rangle$}
        Invoke \gts (Algorithm~\ref{alg:gts}) to get the route $\Route$ of $\tgps$\;
        \ForEach{GPS point $\pii$ in $\tgps$ with segment $e_i$ in $\Route$}{
            $\pis.e\gets e_i$; 
            $\pis.t\gets\pii.t$\;
            Get $\pis.r$ by orthogonal projection of $\pii$ to $e_i$\;
        } 
        $\HM\gets$ Invoke \enc  with input $\tgps$, all $\pis$ of $\pii$ in $\tgps$, route $\Route$ (Eq.\eqref{eq:gseq-token}-\eqref{eq:getHM})\;  
        $\hv_0\gets$mean pooling over $\HM$\;
        $\trajhigh\gets\langle\pseg_1\rangle$\;
        \ForEach{$\pis$ of $\pii$ in $\tgps$}{
        $n_i\gets \lfloor(\pseg_{i+1}.t-\pis.t)/\eps\rfloor$\; 
        \For{$j=1,...,n_i$}{
        Invoke decoding process with $\hj$ and $\HM$ as input of Eq.\eqref{eq:prob-seg-a} to get $w_{k,j}$ for segment $e_k\in  \Route[\ps{j-1}.e,:]$  \; 
        $\forall e_k\in  \Route[\ps{j-1}.e,:]$, get $P(e_k|\pjs)$ by Eq.\eqref{eq:probekaj}\;
        $\pjs.e\gets\arg\max_{e_k\in \Route[\ps{j-1}.e,:] }P(e_k|\pjs)$ by Eq.\eqref{eq:predpjse}\;
        $\pjs.r\gets$ Eq.\eqref{eq:pred-ratio}\;
        $\trajhigh.$append$(\pjs)$\;
        }
        $\trajhigh.$append$(\pseg_{i+1})$\;
        }
    \Return $\trajhigh$\;
\end{algorithm}

Meanwhile, we encode the sequence of segments in route $\Route$ via another transformer.
Specifically, for all $\lR$ segments $e_j$ in $\Route$, we apply an FC layer to get $\Rb \in \mathbb{R}^{\lR \times d_h}$, where the $j$-th row is an embedding of the $j$-th segment $e_j$ in $\Route$. We then adopt a transformer over $\Rb$ to get $\RM \in \mathbb{R}^{\lR \times d_h}$, preserving the sequential patterns in $\Route$, 
\begin{equation}\label{eq:seg-token} 
    \RM=\text{Trans}_\Route(\Rb), \text{ } \Rb=\mathbf{1}_{\Route} \mathbf{W}_7+\mathbf{b}_7, 
\end{equation}
where $\mathbf{1}_{\Route} \in \mathbb{R}^{\lR \times n}$ are the one-hot vectors of the segments in $\Route$, $\mathbf{W}_7\in \mathbb{R}^{n \times d_h}$, $\mathbf{b}_7\in\mathbb{R}^{d_h}$ are the learnable parameters.

Now the question is how to fuse $\TM$ and $\RM$ together to obtain the final representation  $\HM\in\mathbb{R}^{\lR\times d_h}$ that preserves not only the sequences $\tgps$ and $\Route$ but also their relationships.
$\HM$ should have $\lR$ rows, each corresponding to a segment in $\Route$, since the \dec only focuses on the  $\lR$ segments in $\Route$ as candidates to recover missing points. 
To this end, we design an attention mechanism to produce $\HM$.
Intuitively, from the view of a segment $e_j$ in route $\Route$, the points in the sparse trajectory $\tgps$ should have distinct influence on it.
Thus, for every segment $e_j$ in $\Route$ with  embedding $\RM[j,:]$, we compute its attention score $\beta_{i,j}$ for every point $\pii$ with embedding $\TM[i,:]$ in Eq.\eqref{eq:route-score}.
\begin{equation} \label{eq:route-score}
\small
  \beta_{i,j}  = \frac{exp(b_{i,j})}{\sum_{k=1}^{\ell} exp(b_{k,j})} , \text{ } b_{i,j}  = \RM[j,:] \cdot \TM[i,:] 
\end{equation}

The representation of the trajectory $\tgps$ w.r.t. segment $e_j$ in $\Route$ is $\sum_{\pii\in\tgps} \beta_{i,j}\TM[i,:]$.
Then the final embedding of $\HM[i,:]$ is
\begin{equation} \label{eq:getHM}
    \textstyle\HM[j,:]=\RM[j,:]+ \sum_{\pii\in\tgps} \beta_{i,j}\TM[i,:],
\end{equation}
where   $\HM \in \mathbb{R}^{|\routepred_\tgps| \times d_h}$ contains the output embeddings.

\header \textbf{\Dec.}
With the embeddings above as input, the decoding process predicts a missing $\pjs$ by  two concurrent tasks: classifying the segment $\pjs.e$ from the candidate segments in $\Route$, and regressing the position ratio $\pjs.r\in[0,1)$.

In the classification task of the \dec, we determine the segment $\pjs.e$ for the missing point $\pjs$. A straightforward approach would be to map the hidden state $\hj$ to a probability vector matching the number of road segments  in the road network $\rnet$, and then apply a mask that sets the probabilities of segments in $\Route$ to 1 and all others to 0, selecting the segment with the highest probability as $\pjs.e$. However, this method is inefficient for large road networks~\cite{RNTrajRec23,MTrajRec21}. Moreover, since sparse trajectories produce routes of varying lengths, we opt for binary classification on the segments within $\Route$ to predict $\pjs.e$, in contrast to existing methods that use multi-classification over a fixed set of $|E|$ segments in $\rnet$.

As shown in \fig~\ref{fig:archi-rec}, we use GRU~\cite{GRU14} as the sequential model to produce the hidden state $\hj$, which will be used to guide the recovery of the missing $\pjs$. The initial hidden state $\hv_0$ is derived by mean pooling the row vectors of $\HM$. 
When predicting $\pjs$, for the $k$-th segment in $\Route$ with $k$-th row $\HM[k,:]$ as its embedding, we concatenate the embedding with the hidden state $\hj$. This concatenated vector is then processed by a two-layer MLP classifier to produce a scalar value $w_{k,j}$.
\begin{equation} \label{eq:prob-seg-a}
        w_{k,j} = ReLU(\text{concat}[\HM[k:], \hj] \mathbf{W}_8 + \mathbf{b}_8) \mathbf{W}_9 + \mathbf{b}_9
\end{equation}
where $\mathbf{W}_8\in \mathbb{R}^{2d_h \times d_h}$, $\mathbf{b}_8\in \mathbb{R}^{d_h}$, $\mathbf{W}_9\in \mathbb{R}^{d_h \times 1}$, $\mathbf{b}_9\in \mathbb{R}^{1}$ are the learnable parameters.

The scalar value $w_{k,j}$ serves two purposes.
First, we can get the probability that segment $e_k$ is the segment of $\pjs$ with hidden state $\hj$ via a sigmoid function over $w_{k,j}$,   
\begin{equation}\label{eq:probekaj}
    P(e_k|\pjs)=sigmoid(w_{k,j})
\end{equation}

As shown in Algorithm~\ref{alg:TR} (Lines 10-15), we recover missing  points in $\trajhigh$ sequentially. The segments of the recovered  points in $\trajhigh$ must adhere the segment order specified in route $\Route$.
In other words, if the missing point $\pjs$ is recovered after $\ps{j'}$, then the segment $\pjs.e$ should not precede $\ps{j'}.e$ in route $\Route$.
Hence, when predicting segment $\pjs.e$, we select the segment with the highest probability $P(e_k|\pjs)$ among all segments in the sub-Route  of $\Route$ starting from $\ps{j-1}.e$, denoted as $\Route[\ps{j-1}.e,:]$, where  $\ps{j-1}$ is the point just recovered before $\pjs$.
\begin{equation}\label{eq:predpjse}
\pjs.e=\arg\max_{e_k\in \Route[\ps{j-1}.e,:] }P(e_k|\pjs)
\end{equation}

The second usage of $w_{k,j}$ is for the regression task to predict the position ratio $\pjs.r$.
We apply softmax over all $w_{k,j}$ values for  $k=1,...,\lR$, where $\lR$ is the sequence length of route $\Route$, to get $\psi_{k,j}$, which is an attention value  between $\HM[k,:]$ and $\hj$.
Let attention vector be $\mathbf{\psi}_j=[\psi_{1,j},...,\psi_{\lR,j}]$. 
We then apply a weighted row sum in $\HM$ with  each row weighted by  $\psi_{k,j}$, to get $\mathbf{\psi}_j\HM$. 
Finally, we concatenate $\hj$ and $\mathbf{\psi}_j\HM$ to go through an MLP with sigmoid to get the predicted ratio $\pjs.r$,
\begin{equation} \label{eq:pred-ratio}
\small 
    \pjs.r = sigmoid(ReLU(\text{concat}[\mathbf{h}_j, \mathbf{\psi}_j \HM] \mathbf{W}_{10} + \mathbf{b}_{10}) \mathbf{W}_{11} + \mathbf{b}_{11}),
\end{equation}
where $\mathbf{W}_{10}\in \mathbb{R}^{2d_h \times d_h}$, $\mathbf{b}_{10}\in \mathbb{R}^{d_h}$, $\mathbf{W}_{11}\in \mathbb{R}^{d_h \times 1}$, $\mathbf{b}_{11}\in \mathbb{R}^{1}$ are the learnable parameters.

Lastly, in the \dec shown in \fig~\ref{fig:archi-rec}, the current hidden state $\hj$, the predicted segment and ratio of $\pjs$ are then used as the input to obtain the next hidden state $\hv_{j+1}$ for recovering the next missing point $\ps{j+1}$, until  $\trajhigh$ is obtained for the input sparse trajectory $\tgps$.

\header
\textbf{Objective.}
During offline training,   $\pjs$ has ground-truth segment $\pjs.e$, and a segment $e_k$ in route $\Route$ has class label $y_{k,j}=1$ if $e_k = \pseg_j.e$, class label $y_{k,j}=0$ otherwise. The training loss of predicting $\pjs.e$ among all segments in $\Route$ is
\begin{equation} \label{eq:loss-seg}
    {\small \begin{aligned}
        L_{seg} = - \sum_{\forall \pseg_j \in \trajhigh}& \sum_{\forall e_k \in \Route} (y_{k,j} \log P(e_k|\pjs) \\ & +\left(1-y_{k,j}\right) \log \left(1-P(e_k|\pjs)\right))
    \end{aligned}}
\end{equation} 

The prediction of position ratio adopts  mean absolute error, 
\begin{equation}
    \label{eq:loss-ratio}
    \textstyle
    L_{r}= \sum_{\forall \pseg_j \in \trajhigh} |\pseg_j.r - \pseg_j.\hat{r}|
\end{equation}
where $\pseg_j.r$ is the  predicted ratio and $\pseg_j.\hat{r}$ is the  ground truth.

The overall loss of \rec is the weighted combination of the two loss functions in Eq.\eqref{eq:loss-seg} and \eqref{eq:loss-ratio}, 
\begin{equation}
        \label{eq:loss-rec}
\textstyle L_{\rec}=\frac{1}{|\trajdata|} \sum_{\forall \trajhigh \in \trajdata} (L_{seg} + \lambda L_{r})
\end{equation}
where $\lambda$ is a hyperparameter  to trade off the two tasks.

\header 
\textbf{Algorithm.}
In Algorithm~\ref{alg:TR}, after obtaining $\pis$ for every $\pii$ in $\tgps$ over $\Route$ (Lines 1-4), we compute the DualFormer encoding $\HM$ by Eq.\eqref{eq:gseq-token}-\eqref{eq:getHM} (Line 5) and get the initial state $\hv_0$ by mean pooling (Line 6).
After initializing $\trajhigh$ at Line 7, for every pair of consecutive points  $\pis$ and $\pseg_{i+1}$, we calculate the number $n_i$ of missing points necessary to maintain an $\eps$-sampling rate based on their time interval and $\eps$ (Lines 8-9).
From Lines 10 to 15, \algo invokes the decoding process to recover $n_i$ missing points $pjs$ by obtaining probability $P(e_k|\pjs)$ for candidate segments $e_k$ (Lines 11-12), selecting the candidate with the highest probability as $\pjs.e$ (Line 13), and predicting position ratio $\pjs.r$ (Line 14).
The recovered $\pjs$ is appended to $\thigh$ at Line 15, and $\pseg_{i+1}$ of GPS point $\pgps_j$ is appended at Line 16.
The final $\thigh$ is returned at Line 17. 

\header 
\textbf{Complexity.}  
The first step of \algo is \gts with $O(\lt^2d+\lt d^2+\lt \log\numseg + \lt^2\Tilde{deg})$ time.
For the encoding and decoding processes in \algo, the time complexity is as follows.
Let $d$ be the dimension of embeddings involved. 
In the encoding process, we run a transformer over $\tgps$ in $O(\lt^2d)$ time, and another transformer over $\Route$ in $O(\lR^2d)$ time, where $\lR$ is the sequence length of $\Route$. We compute attention between $\tgps$ and $\Route$ in $O(\lt\lR d)$, and the time of MLPs is $O(d^2)$. 
In the decoding process of \rec, we need to recover at most $\lthigh$ points of $\thigh$, which costs $O(\lthigh(\lR+d^2))$.
The time of encoding and decoding in \rec is  $O(\lt^2d+\lR^2d+\lt\lR d+\lthigh(\lR+d^2))$, which simplifies to $O(\lt^2d+\lt d^2)$ if using $\lt$ to represent $\lt,\lR,\lthigh$.
Thus, the overall time complexity of \algo is $O(\lt^2d+\lt d^2+\lt \log\numseg + \lt^2\Tilde{deg})$.

\section{Experiments}\label{sec:exp}

After the experimental setup in Section~\ref{subsec:setup}, we evaluate the efficiency and effectiveness of trajectory recovery in Section~\ref{subsec:effectiveandEfficiencyTR} followed by experimental analysis in Section~\ref{subsec:ExpanalysisTR}. Then we conduct efficiency and effectiveness evaluations and experimental analysis of map matching in Section~\ref{sec:effectivenessandEfficiencyMM}.

\subsection{Experimental Setup} \label{subsec:setup}

\noindent\textbf{Datasets.}
\tab~\ref{tab:datasets} lists the statistics of the  4 real-world trajectory data, including taxi data on Porto District in Portugal (PT)~\cite{Porto} and Beijing (BJ)~\cite{DITA18}, and {DiDi ride-sharing data on Xi'an (XA) and Chengdu (CD) that have been widely used in the literature~\cite{RNTrajRec23,STGED24,TrajCL23}.
All datasets can be downloaded from our code repository.}
All datasets are large with millions of trajectories.
\tab~\ref{tab:datasets} shows the $\eps$-sampling rate in seconds, number of GPS points, length, and travel time of trajectories.
The time period, the number of segments and intersections  and the area of road networks are also provided.
Note that the areas of all datasets already cover the major urban regions of cities,  comparable to or larger than existing studies~\cite{RNTrajRec23}. 
Road networks are from OpenStreetMap~\cite{OpenStreetMap}. 

Note that the $\eps$ sampling rate of each dataset in \tab~\ref{tab:datasets} is regarded as the target high-sampling rate. 
For an $\eps$-sampling trajectory, we generate its sparse trajectory   by randomly sampling the points in it, so that the resulting sparse trajectory $\tgps$ has average interval $\eps/\samp$, where $\samp$ is a ratio in $(0,1)$ to control the level of sparsity. 
By default, $\samp$ is 0.1, meaning that the time interval of sparse trajectories is 10 times larger than the high-sampling trajectories, \ie 150s on PT, 120s on XA, 600s on BJ, and 120s on CD.  
We also vary $\samp$ from 0.1 to 0.5 to study the effect under different sparsity levels.
{To get ground truth, for a raw high-sampling GPS trajectory with $\eps$ sample rate, we use map matching from~\cite{fmm2018} to align it with the road network, to get ground-truth route. For each GPS point 
$p_i$ in the raw trajectory, we draw a perpendicular line from  
$p_i$  to its ground-truth segment. The intersection point is the map-matched point 
$a_i$  for $p_i$, thus obtaining the ground-truth map-matched $\eps$-sampling trajectory $\thighpred$.}
For a dataset, we randomly split its trajectories into training, validation, testing sets with ratio in 40\%, 30\%, and 30\%.

\header 
\textbf{Competitors.} 
For \textit{trajectory recovery}, we compare \algo with baselines in 3 categories. (i)
Trajectory recovery on road networks: \mtrajrec~\cite{MTrajRec21}, \rntrajrec~\cite{RNTrajRec23} and \stged~\cite{STGED24}; \linear that first maps a sparse trajectory to road network by FMM~\cite{fmm2018} and then applies linear interpolation.  
(ii) Free-space trajectory recovery methods \dhtr~\cite{DHTR21} and \teri~\cite{TERI23} that are   extended to road networks by substituting their grid structures with road segments.
(iii) Trajectory representation learning methods followed by a decoder to recover trajectories, including \trajgat~\cite{TrajGAT22}, \trajcl~\cite{TrajCL23}, and \stvec~\cite{STVec22}. 
We follow~\cite{RNTrajRec23,STGED24} to compare this category of methods, and use the decoder in~\cite{MTrajRec21}.
For \textit{map matching}, we compare \gts with the latest methods \lhmm~\cite{LHMM24ICDE} and \graphmm~\cite{GraphMM24TKDE}, \fmm~\cite{fmm2018}, \deepmm~\cite{DeepMM22TMC},  \rntrajrec modified to only return routes, {and \nn that maps a GPS point to its nearest segment.}

\begin{table}[!t]
	\small
	\centering
	\caption{Dataset Statistics }
	\vspace{-1mm} 
	\setlength\tabcolsep{2pt}
	\resizebox{0.9\columnwidth}{!}{%
		\begin{tabular}{lllll}
			\hline
			 & \begin{tabular}[c]{@{}c@{}}Porto\\ (PT)\end{tabular} 
			 & \begin{tabular}[c]{@{}c@{}}Xi'an\\ (XA)\end{tabular}
			 & \begin{tabular}[c]{@{}c@{}}Beijing\\ (BJ)\end{tabular}
			 & \begin{tabular}[c]{@{}c@{}}Chengdu\\ (CD)\end{tabular}   \\
			 \hline
                \# of trajectories & 1,013,437     & 1,426,950  & 1,176,097   & 2,382,422 \\
			$\eps$ sampling rate (s) &15 &12 &60 & 12 \\ 
			Avg \# of points  & 40.21 & 69.36 & 31.59 & 54.32  \\		
                Avg length (\textit{m})   & 4,180.41 & 5,049.27 & 6494.78 & 4,397.41 \\
			Avg travel time ($s$)  & 585.12   & 816.44   & 845.95   & 636.37 \\
			Time period 
			& \begin{tabular}[c]{@{}c@{}}2013/7/1-\\ 2014/6/30\end{tabular} 
			& \begin{tabular}[c]{@{}c@{}}2016/10/1-\\ 2016/10/31\end{tabular}
			& \begin{tabular}[c]{@{}c@{}}2009/3/2-\\ 2009/3/25\end{tabular}
			& \begin{tabular}[c]{@{}c@{}}2016/10/1-\\ 2016/10/31\end{tabular} 
			\\
			rea ($km^2$) & $11.7 \times 5.2$ & $9.1 \times 8.5$ & $29.6 \times 30.0$ & $10.4 \times 10.8$ \\
			\# of segments      & 11,491   & 5,699 & 65,276 & 9,255 \\
                \# of intersections & 5,330    & 2,579 & 28,738 & 3,973 \\
			\hline
		\end{tabular}%
	}
	\label{tab:datasets}
	\vspace{-3mm}
\end{table}

\begin{table*}[!t]
\small
\centering
\caption{Effectiveness of Trajectory Recovery. \textit{Higher} recall, precision, F1, and accuracy (in percentage), and \textit{lower} MAE and RMSE (in meters) represent better performance. Best is in bold and runner-up is underlined.} 
\vspace{-2mm}

    \renewcommand{\arraystretch}{0.97}

\label{tab:comparison_baselines}
\resizebox{0.94\textwidth}{!}{%
\begin{threeparttable}

\begin{tabular}{l|cccccc|cccccc}
    \hline
    & \multicolumn{6}{c|}{PT}& \multicolumn{6}{c}{XA}     \\
    Methods & Recall  & Precision   & F1    & Accuracy & MAE   & RMSE & Recall  & Precision   & F1    & Accuracy & MAE   & RMSE   \\
    \hline
    \linear & 66.42 & 65.85 & 65.83 & 39.54 & 127.6 & 170.1    & 85.65 & 86.58 & 85.73 & 66.26 & 94.2 & 127.1    \\
    \dhtr & \third{69.84} & 73.96 & 71.52 & 47.92 & 135.4 & 181.7    & \third{85.91} & 91.92 & 88.47 & 69.39 & 162.2 & 211.2   \\
    \teri & 67.76 & 72.11 & 69.35 & 43.23 & 180.5 & 249.6    & 83.32 & 90.59 & 86.15 & 60.73 & 222.5 & 301.2   \\
    \trajgat & 56.44 & 74.21 & 63.45 & 39.83 & 188.6 & 251.8    & 75.06 & 88.78 & 80.25 & 60.37 & 203.3 & 265.1   \\
    \trajcl & 60.11 & 77.61 & 67.18 & 43.67 & 152.2 & 204.8    & 75.76 & 89.01 & 80.99 & 62.56 & 154.9 & 204.4   \\
    \stvec & 61.49 & 76.99 & 67.80 & 43.59 & 149.1 & 200.1    & 76.38 & 87.58 & 80.69 & 62.35 & 158.1 & 207.7   \\
    \mtrajrec & 66.24 & 77.33 & 70.93 & 49.72 & 112.1 & 151.5    & 82.58 & 92.18 & 86.65 & 71.19 & 105.9 & 140.3   \\
    \stged & 67.52 & 78.54 & 72.19 & 50.19 & 112.9 & 153.8  & 84.01 & 93.26 & 87.94 & 73.69 & 98.4 & 132.8    \\
    \rntrajrec & 67.29 & \second{79.52} & \third{72.48} & \third{52.22} & \third{102.6} & \third{140.6}   & 84.73 & \third{93.76} & \third{88.61} & \third{74.79} & \third{93.1} & \third{126.5}    \\
    \algo & \first{72.07} & \first{80.92} & \first{75.87} & \first{57.83} & \first{84.10} & \first{121.8}    & \first{86.89} & \first{95.09} & \first{90.44} & \first{78.95} & \first{68.1} & \first{103.1}    \\
    
    \hline
    & \multicolumn{6}{c|}{BJ}     & \multicolumn{6}{c}{CD}       \\
    Methods & Recall  & Precision   & F1    & Accuracy & MAE   & RMSE & Recall  & Precision   & F1    & Accuracy & MAE   & RMSE \\
    \hline
    \linear & 50.28 & 54.13 & 51.54 & 37.35 & 325.5 & 431.3    & 82.66 & 81.82 & 81.77 & 58.17 & 106.2 & 141.5 \\
    \dhtr & 54.41 & 59.61 & 56.16 & 43.77 & 486.7 & 629.4      & 83.14 & 87.22 & 84.68 & 63.84 & 168.3 & 222.3  \\
    \teri & \third{56.61} & 59.34 & 57.23 & 44.34 & 451.5 & 592.1      & 81.62 & 86.07 & 83.15 & 57.99 & 216.6 & 294.7 \\
    \trajgat & 47.95 & 58.64 & 51.29 & 39.41 & 476.5 & 605.4   & 74.42 & 87.56 & 80.05 & 57.95 & 200.4 & 264.2 \\
    \trajcl & 52.63 & \third{64.39} & 57.02 & 43.04 & 397.1 & 509.2    & 75.12 & 87.79 & 80.11 & 60.14 & 152.6 & 204.3  \\
    \stvec & 51.36 & 62.98 & 55.67 & 41.89 & 423.5 & 543.3     & 75.46 & 88.18 & 80.49 & 60.43 & 155.1 & 206.9  \\
    \mtrajrec & 53.35 & 62.44 & 56.68 & 43.58 & 375.1 & 477.2  & 83.34 & 91.24 & 86.65 & 68.42 & 104.8 & 141.1  \\
    \stged & 55.49 & 62.98 & 58.19 & 45.21 & 415.4 & 551.3     & 83.81 & 92.01 & 87.25 & 69.78   & 103.1 & 140.5  \\
    \rntrajrec & 55.39 & 64.38 & \third{58.78} & \third{46.22} & \third{318.2} & \third{413.7} & \third{84.17} & \second{93.26} & \third{88.05} & \third{71.78}   & \third{95.1} & \third{131.8} \\
    \algo & \first{62.15} & \first{66.53} & \first{63.62} & \first{53.71} & \first{234.3} & \first{327.1}    & \first{85.86} & \first{93.95} & \first{89.29} & \first{75.28} & \first{75.1} & \first{114.7} \\
    \hline
\end{tabular}%
\end{threeparttable}

}
	\vspace{-4mm}

\end{table*}

\header  \textbf{Implementations.} 
In our methods, $\candsize$ is set to $10$ as analyzed in Section~\ref{sec:gtsconcepts}. We set dimension $d_0=64$ in Eq.\eqref{eq:fccand}, $d_2=64$ in Eq.\eqref{eq:mlpcand} for candidate segment embedding and point embedding in \gts. We stack two layers and four heads of transformer in Eq.\eqref{eq:te-point}.  
As for the hidden dimension of MLP in Eq.\eqref{eq:mlpcand} and \eqref{eq:attn}, we set $d_1=128$ and $d_3=256$ respectively. 
We set the input dimension of transformer $d_h=64$ in Eq.\eqref{eq:gseq-token} and \eqref{eq:seg-token}, and the FFN dimension in transformer is $512$. 
We stack four DualFormer layers with four heads in \rec.
The learning rate and batch size are 1e-3 and 512 respectively.
Our methods are implemented in Python 3.8 with PyTorch 1.13.
Source codes of all competitors are in Python and  obtained from the respective authors.  We follow their suggested settings to tune optimal parameters.
All methods are trained to converge.
For any method requiring route planning as a subroutine, we adopt the DA-based route in~\cite{DRPK23} for fair comparison.
{Existing studies on route planning~\cite{NMLR21,DRPK23} indicate that it is possible but rare for the output route from a source $e_i$ to fail in reaching destination $e_{i+1}$, with a very low probability, such as 0.06\% on PT in experiments. This has a negligible impact on the overall map-matching quality. In such a case, a practical solution is to use the fastest route to connect $e_i$ and $e_{i+1}$.} 
All experiments are conducted on a Linux machine powered by Intel Xeon® Gold 6226R 2.90GHz CPU and  NVIDIA GTX 3090 GPU with 24GB video memory.

\header \textbf{Evaluation Metrics.}  (i) \textit{Efficiency metrics}. 
We evaluate a method by  inference time to recover trajectories and training time per epoch.
(ii) \textit{Effectiveness metrics}.
For trajectory recovery of   $\tgps=\langle \pgps_1,\pgps_2,...\pgps_\lt\rangle$ with map-matched $\eps$-sampling trajectory $\trajhigh=\langle \pseg_1,\pseg_2,...,\pseg_{\lthigh}\rangle$,  
we  follow~\cite{MTrajRec21,RNTrajRec23} to use Mean Absolute Error (MAE) and Root Mean Square Error (RMSE)  to evaluate the distance error of the recovered map-matched points over  ground truth, 
\begin{equation}\label{eq:metrics-dist}
\scriptsize   
     {MAE}(\thigh, \thighpred) = \frac{1}{\lthigh} \sum_{i=1}^{\lthigh} |d(\pseg_i, \pspred_i)|;   {RMSE}(\thigh, \thighpred) = \sqrt{\frac{1}{\lthigh} \sum_{i=1}^{\lthigh} (d(\pseg_i, \pspred_i))^2},
\end{equation}
where $d(\pseg_i, \hat{\pseg}_i)$ is the road network distance between predicted $\pseg_i$ and ground truth $\hat{\pseg}_i$.

To evaluate if a  method recovers the segments  of a trajectory, we adopt  4 popular metrics~\cite{RNTrajRec23,MTrajRec21}: Precision, Recall, F1 score and  Accuracy.
For \textit{trajectory recovery}, let $S$ represent the segments of all $\pis$ in  $\thigh$ and $\hat{S}$ represent the segments in the ground-truth $\thighpred$, and the 4 metrics are calculated as follows.
When evaluating \textit{map matching}, $S$ becomes the returned route $\Route$ while $\hat{S}$ is ground-truth route; following existing methods~\cite{LHMM24ICDE,DMM24TKDE}, 
we use Recall, Precision, and F1, as well as Jaccard similarity $Jaccard(S,\hat{S})=\frac{|S \cap \hat{S}|}{|S\cup\hat{S}|}$.
\begin{equation*}
    \label{eq:metrics}
    \footnotesize
    \vspace{-1mm}
    \begin{aligned}
        & Recall(S, \hat{S}) = \frac{|S \cap \hat{S}|}{|S|}
        & F1(S, \hat{S}) = \frac{2 pre(S, \hat{S})  rec(S, \hat{S})}{pre(S, \hat{S}) + rec(S, \hat{S})} \\
        & Precision(S, \hat{S}) = \frac{|S \cap \hat{S}|}{|\hat{S}|}
        & Accuracy(S, \hat{S}) = \frac{1}{\lthigh} \sum_{i=1}^{\lthigh} \textbf{1}\{s_i = \hat{s}_i\}
    \end{aligned}
\end{equation*}

For each metric, we calculate the metric score per trajectory and report the average over all testing trajectories.

\pgfplotsset{ every non boxed x axis/.append style={x axis line style=-} }
\pgfplotsset{ every non boxed y axis/.append style={y axis line style=-} }
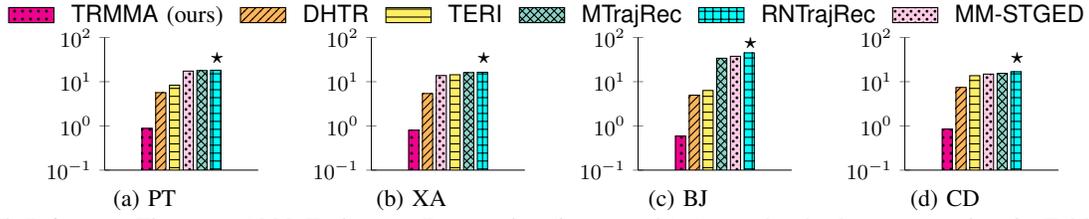
\begin{figure*}[!t]
\centering
\begin{small}
\begin{tikzpicture}[scale=0.9]
    \begin{customlegend}[
        legend entries={\algo (ours),\dhtr,\teri,\mtrajrec,\rntrajrec,\stged},
        legend columns=6,
        area legend,
        legend style={at={(0.45,1.15)},anchor=north,draw=none,font=\small,column sep=0.15cm}]
        
        \addlegendimage{preaction={fill, magenta}, pattern=dots}
        \addlegendimage{preaction={fill, myorange}, pattern={north east lines}}
        \addlegendimage{preaction={fill, myyellow},pattern={horizontal lines}}
        \addlegendimage{preaction={fill, mycyan},pattern=crosshatch}
        \addlegendimage{preaction={fill, mycyan2}, pattern=grid}
        \addlegendimage{preaction={fill, mypink},pattern=crosshatch dots}
    \end{customlegend}

\end{tikzpicture}
\\[-\lineskip]
\vspace{-2mm}
\captionsetup{margin={-0.5cm,0cm}}
\subfloat[PT]{
\begin{tikzpicture}[scale=0.9]
\begin{axis}[
    height=\columnwidth/2.5,
    width=\columnwidth/2.3,
    axis lines=left,
    xtick=\empty,
    ybar=1.5pt,
    bar width=0.15cm,
    enlarge x limits=true,
    ymin=0.1,
    ymax=100,
    ytick={0.1,1,10,100,1000,10000},
    ymode=log,
    log origin y=infty,
    log basis y={10},
    every axis y label/.style={at={(current axis.north west)},right=10mm,above=0mm},
    legend style={at={(0.02,0.98)},anchor=north west,cells={anchor=west},font=\tiny}
    ]
        \addplot[preaction={fill, magenta}, pattern=dots] coordinates {(1,0.88)};
        \addplot[preaction={fill, myorange}, pattern={north east lines}] coordinates {(1,5.65)};
        \addplot[preaction={fill, myyellow},pattern={horizontal lines}] coordinates {(1,8.35)};
        \addplot[preaction={fill, mypink},pattern=crosshatch dots] coordinates {(1,17.36)};
        \addplot[preaction={fill, mycyan},pattern=crosshatch]coordinates {(1,17.85)};
        \addplot[preaction={fill, mycyan2}, pattern=grid] coordinates {(1,18.17)};
        \node[above,black] at (axis cs:1.093,18){$\boldsymbol{\star}$};

\end{axis}
\end{tikzpicture}\vspace{-2mm}\hspace{2mm}\label{fig:infer_rec_pt}%
}\quad
\subfloat[XA]{
\begin{tikzpicture}[scale=0.9]
\begin{axis}[
    height=\columnwidth/2.5,
    width=\columnwidth/2.3,
    axis lines=left,
    xtick=\empty,
    ybar=1.5pt,
    bar width=0.15cm,
    enlarge x limits=true,
    xticklabel=\empty,
    ymin=0.1,
    ymax=100,
    ytick={0.1,1,10,100,1000,10000},
    ymode=log,
    log origin y=infty,
    log basis y={10},
    every axis y label/.style={at={(current axis.north west)},right=10mm,above=0mm},
    legend style={at={(0.02,0.98)},anchor=north west,cells={anchor=west},font=\tiny}
    ]
        \addplot[preaction={fill, magenta}, pattern=dots] coordinates {(1,0.80)};
        \addplot[preaction={fill, myorange}, pattern={north east lines}] coordinates {(1,5.39)};
        \addplot[preaction={fill, mypink},pattern=crosshatch dots] coordinates {(1,13.77)};
        \addplot[preaction={fill, myyellow},pattern={horizontal lines}] coordinates {(1,14.34)};
        \addplot[preaction={fill, mycyan},pattern=crosshatch]coordinates {(1,16.03)};
        \addplot[preaction={fill, mycyan2}, pattern=grid] coordinates {(1,16.35)};
        \node[above,black] at (axis cs:1.093,18){$\boldsymbol{\star}$};
\end{axis}
\end{tikzpicture}\vspace{-2mm}\hspace{2mm}\label{fig:infer_rec_xa}%
}\quad
\subfloat[BJ]{
\begin{tikzpicture}[scale=0.9]
\begin{axis}[
    height=\columnwidth/2.5,
    width=\columnwidth/2.3,
    axis lines=left,
    xtick=\empty,
    ybar=1.5pt,
    bar width=0.15cm,
    enlarge x limits=true,
    xticklabel=\empty,
    ymin=0.1,
    ymax=100,
    ytick={0.1,1,10,100,1000,10000},
    ymode=log,
    log origin y=infty,
    log basis y={10},
    every axis y label/.style={at={(current axis.north west)},right=10mm,above=0mm},
    legend style={at={(0.02,0.98)},anchor=north west,cells={anchor=west},font=\tiny}
    ]
        \addplot[preaction={fill, magenta}, pattern=dots] coordinates {(1,0.59)};
        \addplot[preaction={fill, myorange}, pattern={north east lines}] coordinates {(1,4.96)};
        \addplot[preaction={fill, myyellow},pattern={horizontal lines}] coordinates {(1,6.33)};
        \addplot[preaction={fill, mycyan},pattern=crosshatch]coordinates {(1,33.91)};
        \addplot[preaction={fill, mypink},pattern=crosshatch dots] coordinates {(1,37.54)};
        \addplot[preaction={fill, mycyan2}, pattern=grid] coordinates {(1,44.79)};
        \node[above,black] at (axis cs:1.093,40){$\boldsymbol{\star}$};
\end{axis}
\end{tikzpicture}\vspace{-2mm}\hspace{2mm}\label{fig:infer_rec_bj}%
}\quad
\subfloat[CD]{
\begin{tikzpicture}[scale=0.9]
\begin{axis}[
    height=\columnwidth/2.5,
    width=\columnwidth/2.3,
    axis lines=left,
    xtick=\empty,
    ybar=1.5pt,
    bar width=0.15cm,
    enlarge x limits=true,
    xticklabel=\empty,
    ymin=0.1,
    ymax=100,
    ytick={0.1,1,10,100,1000,10000},
    ymode=log,
    log origin y=infty,
    log basis y={10},
    every axis y label/.style={at={(current axis.north west)},right=10mm,above=0mm},
    legend style={at={(0.02,0.98)},anchor=north west,cells={anchor=west},font=\tiny}
    ]
        \addplot[preaction={fill, magenta}, pattern=dots] coordinates {(1,0.85)};
        \addplot[preaction={fill, myorange}, pattern={north east lines}] coordinates {(1,7.41)};
        \addplot[preaction={fill, myyellow},pattern={horizontal lines}] coordinates {(1,13.68)};
        \addplot[preaction={fill, mypink},pattern=crosshatch dots] coordinates {(1,14.84)};
        \addplot[preaction={fill, mycyan},pattern=crosshatch]coordinates {(1,15.44)};
        \addplot[preaction={fill, mycyan2}, pattern=grid] coordinates {(1,16.81)};
        \node[above,black] at (axis cs:1.093,18){$\boldsymbol{\star}$};
\end{axis}
\end{tikzpicture}\vspace{-2mm}\hspace{2mm}\label{fig:infer_rec_cd}%
}%
\vspace{-2mm}
\end{small}
\caption{Inference Time per 1000 Trajectory Recoveries (in seconds) ($\boldsymbol{\star}$ marks the best competitor in \tab \ref{tab:comparison_baselines}).} 
\label{fig:infer_rec}
\vspace{-3mm}
\end{figure*}

\subsection{Effectiveness and Efficiency of Trajectory Recovery} \label{subsec:effectiveandEfficiencyTR}

\noindent \textbf{Effectiveness.}
\tab~\ref{tab:comparison_baselines} reports the recall, precision, F1, accuracy, MAE, and RMSE of all methods for trajectory recovery. An overall observation is that our method \algo consistently achieves the best performance on all datasets under all evaluation metrics, outperforming existing methods often by a significant margin.
For instance, 
On PT, \algo achieves 75.87\% F1 score, while the F1 of \rntrajrec is 72.48\%, indicating 3.39\% improvement. Further, on PT, \rntrajrec achieves strong precision at the cost of low recall, while  \algo can improve both metrics.
on BJ, \algo achieves 62.15\% recall score, indicating 5.54\% improvement over the best competitor \teri with 56.61\% recall.
On XA, \algo has MAE  68.1\textit{m} and RMSE 103.1\textit{m}, achieving 26.85\% and 18.5\%  relative improvements over \rntrajrec with  93.1\textit{m} and 126.5\textit{m} respectively. A similar observation can be made on BJ, where \algo has MAE 234.3\textit{m}, significantly lower than the MAE of \rntrajrec 318.2\textit{m}. 
\tab~\ref{tab:comparison_baselines} demonstrates that \algo is effective in accurately predicting the correct segments and position ratios to infer missing points to get $\thigh$. This is achieved by the proposed techniques in \algo, including the encoding and decoding modules in Section~\ref{sec:rec}, and the map matching method \gts in Section~\ref{sec:gps2seg}, which will be evaluated independently in Section~\ref{sec:effectivenessandEfficiencyMM}. 
Additionally, 
\linear and other baselines in the category of trajectory representation learning, including \trajgat, \trajcl, and \stvec,  achieve inferior performance, indicating  the importance of tailoring techniques for trajectory recovery.
Due to space limit, in what follows, we omit them.

\pgfplotsset{ every non boxed x axis/.append style={x axis line style=-} }
\pgfplotsset{ every non boxed y axis/.append style={y axis line style=-} }
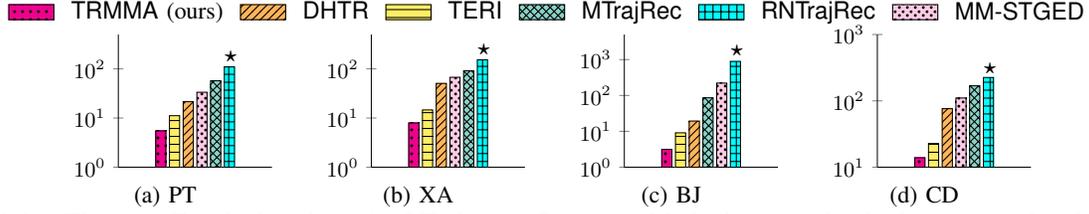
\begin{figure*}[!t]
\centering
\begin{small}
\begin{tikzpicture}[scale=0.9]
    \begin{customlegend}[
        legend entries={\algo (ours),\dhtr,\teri,\mtrajrec,\rntrajrec,\stged},
        legend columns=6,
        area legend,
        legend style={at={(0.45,1.15)},anchor=north,draw=none,font=\small,column sep=0.15cm}]
        
        \addlegendimage{preaction={fill, magenta}, pattern=dots}
        \addlegendimage{preaction={fill, myorange}, pattern={north east lines}}
        \addlegendimage{preaction={fill, myyellow},pattern={horizontal lines}}
        \addlegendimage{preaction={fill, mycyan},pattern=crosshatch}
        \addlegendimage{preaction={fill, mycyan2}, pattern=grid}
        \addlegendimage{preaction={fill, mypink},pattern=crosshatch dots}
    \end{customlegend}

\end{tikzpicture}
\\[-\lineskip]
\vspace{-2mm}
\captionsetup{margin={-0.5cm,0cm}}
\subfloat[PT]{
\begin{tikzpicture}[scale=0.9]
\begin{axis}[
    height=\columnwidth/2.5,
    width=\columnwidth/2.3,
    axis lines=left,
    xtick=\empty,
    ybar=1.5pt,
    bar width=0.15cm,
    enlarge x limits=true,
    ymin=1,
    ymax=500,
    ytick={0.1,1,10,100,1000,10000},
    ymode=log,
    log origin y=infty,
    log basis y={10},
    every axis y label/.style={at={(current axis.north west)},right=10mm,above=0mm},
    legend style={at={(0.02,0.98)},anchor=north west,cells={anchor=west},font=\tiny}
    ]
        \addplot[preaction={fill, magenta}, pattern=dots] coordinates {(1,329.5/60)};
        \addplot[preaction={fill, myyellow},pattern={horizontal lines}] coordinates {(1,669.8/60)};
        \addplot[preaction={fill, myorange}, pattern={north east lines}] coordinates {(1,1289.5/60)};
        \addplot[preaction={fill, mypink},pattern=crosshatch dots] coordinates {(1,2007.4/60)};
        \addplot[preaction={fill, mycyan},pattern=crosshatch]coordinates {(1,3436.9/60)};
        \addplot[preaction={fill, mycyan2}, pattern=grid] coordinates {(1,6582.1/60)};
        \node[above,black] at (axis cs:1.093,100){$\boldsymbol{\star}$};

\end{axis}
\end{tikzpicture}\vspace{-2mm}\hspace{2mm}\label{fig:train_rec_pt}%
}\quad
\subfloat[XA]{
\begin{tikzpicture}[scale=0.9]
\begin{axis}[
    height=\columnwidth/2.5,
    width=\columnwidth/2.3,
    axis lines=left,
    xtick=\empty,
    ybar=1.5pt,
    bar width=0.15cm,
    enlarge x limits=true,
    xticklabel=\empty,
    ymin=1,
    ymax=500,
    ytick={0.1,1,10,100,1000,10000},
    ymode=log,
    log origin y=infty,
    log basis y={10},
    every axis y label/.style={at={(current axis.north west)},right=10mm,above=0mm},
    legend style={at={(0.02,0.98)},anchor=north west,cells={anchor=west},font=\tiny}
    ]
        \addplot[preaction={fill, magenta}, pattern=dots] coordinates {(1,477.1/60)};
        \addplot[preaction={fill, myyellow},pattern={horizontal lines}] coordinates {(1,876.4/60)};
        \addplot[preaction={fill, myorange}, pattern={north east lines}] coordinates {(1,3038.4/60)};
        \addplot[preaction={fill, mypink},pattern=crosshatch dots] coordinates {(1,4045.5/60)};
        \addplot[preaction={fill, mycyan},pattern=crosshatch]coordinates {(1,5487.1/60)};
        \addplot[preaction={fill, mycyan2}, pattern=grid] coordinates {(1,9177.5/60)};
        \node[above,black] at (axis cs:1.093,140){$\boldsymbol{\star}$};
\end{axis}
\end{tikzpicture}\vspace{-2mm}\hspace{2mm}\label{fig:train_rec_xa}%
}\quad
\subfloat[BJ]{
\begin{tikzpicture}[scale=0.9]
\begin{axis}[
    height=\columnwidth/2.5,
    width=\columnwidth/2.3,
    axis lines=left,
    xtick=\empty,
    ybar=1.5pt,
    bar width=0.15cm,
    enlarge x limits=true,
    xticklabel=\empty,
    ymin=1,
    ymax=5000,
    ytick={0.1,1,10,100,1000,10000},
    ymode=log,
    log origin y=infty,
    log basis y={10},
    every axis y label/.style={at={(current axis.north west)},right=10mm,above=0mm},
    legend style={at={(0.02,0.98)},anchor=north west,cells={anchor=west},font=\tiny}
    ]
        \addplot[preaction={fill, magenta}, pattern=dots] coordinates {(1,188.1/60)};
        \addplot[preaction={fill, myyellow},pattern={horizontal lines}] coordinates {(1,544.7/60)};
        \addplot[preaction={fill, myorange}, pattern={north east lines}] coordinates {(1,1158.6/60)};
        \addplot[preaction={fill, mycyan},pattern=crosshatch]coordinates {(1,5165.2/60)};
        \addplot[preaction={fill, mypink},pattern=crosshatch dots] coordinates {(1,13260.6/60)};
        \addplot[preaction={fill, mycyan2}, pattern=grid] coordinates {(1,53345.6/60)};
        \node[above,black] at (axis cs:1.093,800){$\boldsymbol{\star}$};
\end{axis}
\end{tikzpicture}\vspace{-2mm}\hspace{2mm}\label{fig:train_rec_bj}%
}\quad
\subfloat[CD]{
\begin{tikzpicture}[scale=0.9]
\begin{axis}[
    height=\columnwidth/2.5,
    width=\columnwidth/2.3,
    axis lines=left,
    xtick=\empty,
    ybar=1.5pt,
    bar width=0.15cm,
    enlarge x limits=true,
    xticklabel=\empty,
    ymin=10,
    ymax=1000,
    ytick={0.1,1,10,100,1000,10000},
    ymode=log,
    log origin y=infty,
    log basis y={10},
    every axis y label/.style={at={(current axis.north west)},right=10mm,above=0mm},
    legend style={at={(0.02,0.98)},anchor=north west,cells={anchor=west},font=\tiny}
    ]
        \addplot[preaction={fill, magenta}, pattern=dots] coordinates {(1,833.5/60)};
        \addplot[preaction={fill, myyellow},pattern={horizontal lines}] coordinates {(1,1364.5/60)};
        \addplot[preaction={fill, myorange}, pattern={north east lines}] coordinates {(1,4574.8/60)};
        \addplot[preaction={fill, mypink},pattern=crosshatch dots] coordinates {(1,6623.2/60)};
        \addplot[preaction={fill, mycyan},pattern=crosshatch]coordinates {(1,10114.9/60)};
        \addplot[preaction={fill, mycyan2}, pattern=grid] coordinates {(1,13491.4/60)};
        \node[above,black] at (axis cs:1.093,200){$\boldsymbol{\star}$};
\end{axis}
\end{tikzpicture}\vspace{-2mm}\hspace{2mm}\label{fig:train_rec_cd}%
}%
\vspace{-2mm}
\end{small}
\caption{Training Time per Epoch (in minutes) of Trajectory Recovery Methods ($\boldsymbol{\star}$ marks the best competitor  in \tab \ref{tab:comparison_baselines}).} 
\label{fig:train_rec}
\vspace{-4mm}
\end{figure*}

\header \textbf{Efficiency.}
\fig~\ref{fig:infer_rec} reports the average \textit{inference time} per 1000 sparse trajectory recoveries in seconds.
\algo is faster than the competitors, and more importantly, compared with the best competitor \rntrajrec that achieves high effectiveness in \tab~\ref{tab:comparison_baselines}, \algo is significantly faster, often by orders of magnitude.
For example, on BJ, \algo only needs 0.59s to recover 1000 trajectories, 75.9$\times$ faster than \rntrajrec that needs 44.79s.
Although \dhtr is relatively quick, it delivers lower quality in \tab~\ref{tab:comparison_baselines}. 
\fig~\ref{fig:train_rec} reports the \textit{training time} per epoch.
\algo is much faster to train than the competitors. 
For instance, on XA dataset, our method \algo only needs 7.95 minutes per epoch to train, 19$\times$ faster than \rntrajrec  that costs 152.95 minutes per epoch. 
\dhtr and \teri are relatively fast to train, but produce inferior quality as reported in \tab~\ref{tab:comparison_baselines}.
The efficiency of \algo stems from the technical designs in Section~\ref{sec:rec} with powerful encoding and decoding processes. \algo first identifies a route $\Route$ of $\tgps$ and then focuses solely on the segments in $\Route$ for recovery. In contrast, existing methods~\cite{RNTrajRec23,MTrajRec21} consider all segments from a road network as candidates. The efficiency of \algo aligns with the time complexity analyzed in Section~\ref{sec:rec}.

\begin{figure}[!t]
\centering
\begin{small}
\resizebox{0.9\columnwidth}{!}{%
\begin{tikzpicture}[scale=0.1]
    \begin{customlegend}[legend columns=4,
	    legend entries={\algo,\dhtr,\teri,\linear,\mtrajrec,\rntrajrec,\stged},
	    legend style={at={(0.45,1.15)},anchor=north,draw=none,font=\tiny,column sep=0.05cm}]
	    
	    \addlegendimage{line width=0.25mm,color=red,mark=triangle*}
 	    \addlegendimage{line width=0.25mm,color=magenta,mark=o}
 	    \addlegendimage{line width=0.25mm,color=orange,mark=diamond}
	    \addlegendimage{line width=0.25mm,color=violet,mark=pentagon}
	    \addlegendimage{line width=0.25mm,color=cyan,mark=triangle}
 	    \addlegendimage{line width=0.25mm,color=green,mark=star}
 	    \addlegendimage{line width=0.25mm,color=blue,mark=+}
    \end{customlegend}
\end{tikzpicture}
}
\\[-\lineskip]
\vspace{-1mm}

\begin{tabular}{cc}

\subcaptionbox{PT}
{
    \vspace{-2mm}
	\begin{tikzpicture}[scale=0.95]
		\begin{axis}[
		height=\columnwidth/2.5,
		width=\columnwidth/2.0,
		ylabel={\em Accuracy},
		xmin=1, xmax=5,
		xtick={1,2,3,4,5},
		xticklabels={0.1,0.2,0.3,0.4,0.5},
		every axis y label/.style={at={(current axis.north west)},right=3mm,above=0mm},
		every axis x label/.style={at={(current axis.south)},above=-7mm},
		label style={font=\scriptsize},
		tick label style={font=\scriptsize},
		]
		
		\addplot[line width=0.25mm,color=violet,mark=pentagon]
		table[x=ratio, y=linear, col sep=comma] {data/ratio_pt.tex};
		
		\addplot[line width=0.25mm,color=magenta,mark=o]
		table[x=ratio, y=dhtr, col sep=comma] {data/ratio_pt.tex};
		
		\addplot[line width=0.25mm,color=cyan,mark=triangle]
		table[x=ratio, y=mtraj, col sep=comma] {data/ratio_pt.tex};
		
		\addplot[line width=0.25mm,color=green,mark=star]
		table[x=ratio, y=rntraj, col sep=comma] {data/ratio_pt.tex};
		
		\addplot[line width=0.25mm,color=orange,mark=diamond]
		table[x=ratio, y=teri, col sep=comma] {data/ratio_pt.tex};
		
		\addplot[line width=0.25mm,color=blue,mark=+]
		table[x=ratio, y=stged, col sep=comma] {data/ratio_pt.tex};
		
		\addplot[line width=0.25mm,color=red,mark=triangle*]
		table[x=ratio, y=ours, col sep=comma] {data/ratio_pt.tex};
		
		\end{axis}
	\end{tikzpicture}
}%
& \subcaptionbox{XA}
{
    \vspace{-2mm}
	\begin{tikzpicture}[scale=0.95]
		\begin{axis}[
		height=\columnwidth/2.5,
		width=\columnwidth/2.0,
		ylabel={\em Accuracy},
		xmin=1, xmax=5,
		xtick={1,2,3,4,5},
		xticklabels={0.1,0.2,0.3,0.4,0.5},
		every axis y label/.style={at={(current axis.north west)},right=3mm,above=0mm},
		every axis x label/.style={at={(current axis.south)},above=-7mm},
		label style={font=\scriptsize},
		tick label style={font=\scriptsize},
		]
		
		\addplot[line width=0.25mm,color=violet,mark=pentagon]
		table[x=ratio, y=linear, col sep=comma] {data/ratio_xa.tex};
		
		\addplot[line width=0.25mm,color=magenta,mark=o]
		table[x=ratio, y=dhtr, col sep=comma] {data/ratio_xa.tex};
		
		\addplot[line width=0.25mm,color=cyan,mark=triangle]
		table[x=ratio, y=mtraj, col sep=comma] {data/ratio_xa.tex};
		
		\addplot[line width=0.25mm,color=green,mark=star]
		table[x=ratio, y=rntraj, col sep=comma] {data/ratio_xa.tex};
		
		\addplot[line width=0.25mm,color=orange,mark=diamond]
		table[x=ratio, y=teri, col sep=comma] {data/ratio_xa.tex};
		
		\addplot[line width=0.25mm,color=blue,mark=+]
		table[x=ratio, y=stged, col sep=comma] {data/ratio_xa.tex};
		
		\addplot[line width=0.25mm,color=red,mark=triangle*]
		table[x=ratio, y=ours, col sep=comma] {data/ratio_xa.tex};
		
		\end{axis}
	\end{tikzpicture}
} \\
\subcaptionbox{BJ}
{
    \vspace{-2mm}
	\begin{tikzpicture}[scale=0.95]
		\begin{axis}[
		height=\columnwidth/2.5,
		width=\columnwidth/2.0,
		ylabel={\em Accuracy},
		xmin=1, xmax=5,
		xtick={1,2,3,4,5},
		xticklabels={0.1,0.2,0.3,0.4,0.5},
		every axis y label/.style={at={(current axis.north west)},right=3mm,above=0mm},
		every axis x label/.style={at={(current axis.south)},above=-7mm},
		label style={font=\scriptsize},
		tick label style={font=\scriptsize},
		]
		
		\addplot[line width=0.25mm,color=violet,mark=pentagon]
		table[x=ratio, y=linear, col sep=comma] {data/ratio_bj.tex};
		
		\addplot[line width=0.25mm,color=magenta,mark=o]
		table[x=ratio, y=dhtr, col sep=comma] {data/ratio_bj.tex};
		
		\addplot[line width=0.25mm,color=cyan,mark=triangle]
		table[x=ratio, y=mtraj, col sep=comma] {data/ratio_bj.tex};
		
		\addplot[line width=0.25mm,color=green,mark=star]
		table[x=ratio, y=rntraj, col sep=comma] {data/ratio_bj.tex};
		
		\addplot[line width=0.25mm,color=orange,mark=diamond]
		table[x=ratio, y=teri, col sep=comma] {data/ratio_bj.tex};
		
		\addplot[line width=0.25mm,color=blue,mark=+]
		table[x=ratio, y=stged, col sep=comma] {data/ratio_bj.tex};
		
		\addplot[line width=0.25mm,color=red,mark=triangle*]
		table[x=ratio, y=ours, col sep=comma] {data/ratio_bj.tex};
		
		\end{axis}
	\end{tikzpicture}
}%
& \subcaptionbox{CD}
{
    \vspace{-2mm}
	\begin{tikzpicture}[scale=0.95]
		\begin{axis}[
		height=\columnwidth/2.5,
		width=\columnwidth/2.0,
		ylabel={\em Accuracy},
		xmin=1, xmax=5,
		xtick={1,2,3,4,5},
		xticklabels={0.1,0.2,0.3,0.4,0.5},
		every axis y label/.style={at={(current axis.north west)},right=3mm,above=0mm},
		every axis x label/.style={at={(current axis.south)},above=-7mm},
		label style={font=\scriptsize},
		tick label style={font=\scriptsize},
		]
		
		\addplot[line width=0.25mm,color=violet,mark=pentagon]
		table[x=ratio, y=linear, col sep=comma] {data/ratio_cd.tex};
		
		\addplot[line width=0.25mm,color=magenta,mark=o]
		table[x=ratio, y=dhtr, col sep=comma] {data/ratio_cd.tex};
		
		\addplot[line width=0.25mm,color=cyan,mark=triangle]
		table[x=ratio, y=mtraj, col sep=comma] {data/ratio_cd.tex};
		
		\addplot[line width=0.25mm,color=green,mark=star]
		table[x=ratio, y=rntraj, col sep=comma] {data/ratio_cd.tex};
		
		\addplot[line width=0.25mm,color=orange,mark=diamond]
		table[x=ratio, y=teri, col sep=comma] {data/ratio_cd.tex};
		
		\addplot[line width=0.25mm,color=blue,mark=+]
		table[x=ratio, y=stged, col sep=comma] {data/ratio_cd.tex};
		
		\addplot[line width=0.25mm,color=red,mark=triangle*]
		table[x=ratio, y=ours, col sep=comma] {data/ratio_cd.tex};
		
		\end{axis}
	\end{tikzpicture}
} \\
\end{tabular}%

\vspace{-2mm}
\end{small}
\caption{Trajectory Recovery with Varied Levels of Sparsity} 
\label{fig:exp-ratio}
\vspace{-3mm}
\end{figure}
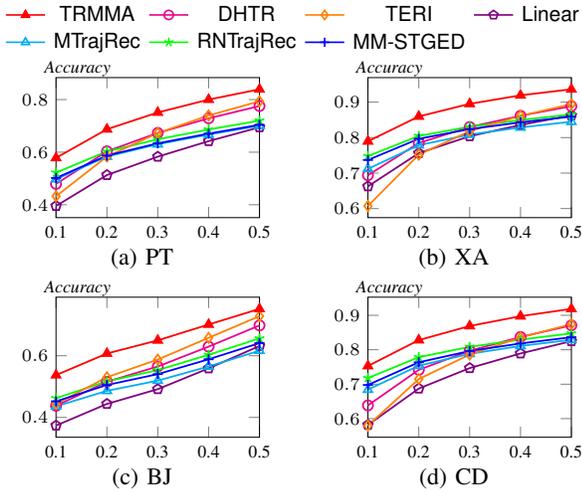

\subsection{Experimental Analysis of Trajectory Recovery} \label{subsec:ExpanalysisTR}

\noindent \textbf{Varying Level of Sparsity.}
As mentioned, given a trajectory with the target $\eps$ sampling rate, we generate its sparse trajectory by random sampling with  time interval $\eps/\samp$, where $\samp$ controls the level of sparsity. We vary $\samp$ from 0.1 to 0.5, and report the accuracy results in \fig~\ref{fig:exp-ratio}.
As trajectories become sparser with a smaller $\samp$, all methods achieve degraded performance. Nevertheless,  the performance gap between \algo and  the competitors maintains under all settings on all datasets. For instance, on PT dataset, when $\samp=0.5$, \algo has 0.8392 accuracy, $4.57\%$ higher than  \teri with 0.7935 accuracy.

\begin{table}[!t]
    \small
    \centering
    \caption{Ablation Result of \algo by Accuracy (\%)}
    \vspace{-2mm}
    \label{tab:ablation}
    
    \renewcommand{\arraystretch}{1}
    \resizebox{0.76\columnwidth}{!}{%
    \setlength{\tabcolsep}{7pt}
        \begin{tabular}{l|cccc}
            \hline
             & PT     & XA     & BJ     & CD     \\
            \hline
            \algo       & \textbf{57.83} & \textbf{78.95} & \textbf{53.71} & \textbf{75.28} \\
            \algo-HMM        & 53.54 & 76.81 & 49.57 & 70.63 \\
            \algo-Near        & 47.01 & 65.81 & 43.66 & 56.22 \\
            {\MMAlinear}  & {43.74} & {68.99} & {41.72} & {62.82} \\
            {\nnlinear}   & {35.45} & {58.03} & {33.97} & {47.61} \\
            \algo-DF     & 54.83 & 77.62 & 50.73 & 73.91 \\
            \algo-C  & 56.85 & 78.63 & 52.13 & 74.96 \\
            \algo-DI & 51.02 & 71.47 & 45.83 & 69.15 \\
            \hline
        \end{tabular}%
    }
    \vspace{-4mm}
\end{table}

\header \textbf{Ablation Study.}
We ablate the techniques in \algo and report the accuracy in \tab~\ref{tab:ablation}.
\algo-Near is \algo without \gts in Section~\ref{sec:gps2seg}, but with nearest segment of a GPS point $\pii$ as its segment $\pis.e$.
\algo-HMM is \algo without \gts, but with HMM~\cite{fmm2018}.
\algo-DF is \algo without the \enc in Section~\ref{sec:rec}, \ie using $\RM$ as $\HM$ in \fig~\ref{fig:archi-rec}.
\algo-C is \algo with \gts that does not consider the candidates in $\Cpi$ as the context to get $\piemb$ in the point embedding module in Section~\ref{sec:gtsArchi}.
\algo-DI is \algo with \gts that does not consider the directional information. 
{\MMAlinear is \gts with linear interpolation  for recovery. \nnlinear is \nn with linear interpolation.}
In \tab~\ref{tab:ablation}, the accuracy of \algo is higher than all ablated versions, validating the effectiveness of our techniques.

\begin{figure}[!t]
\centering
\begin{small}
\resizebox{0.9\columnwidth}{!}{%
\begin{tikzpicture}
    \begin{customlegend}[legend columns=4,
	    legend entries={\algo,\dhtr,\teri,\linear,\mtrajrec,\rntrajrec,\stged},
	    legend style={at={(0.45,1.15)},anchor=north,draw=none,font=\scriptsize,column sep=0.05cm}]
	    
	    \addlegendimage{line width=0.25mm,color=red,mark=triangle*}
 	    \addlegendimage{line width=0.25mm,color=magenta,mark=o}
 	    \addlegendimage{line width=0.25mm,color=orange,mark=diamond}
	    \addlegendimage{dashed,color=red}
	    \addlegendimage{line width=0.25mm,color=cyan,mark=triangle}
 	    \addlegendimage{line width=0.25mm,color=green,mark=star}
 	    \addlegendimage{line width=0.25mm,color=blue,mark=+}
    \end{customlegend}
\end{tikzpicture}
}
\\[-\lineskip]
\vspace{-1mm}

\begin{tabular}{cc}

\subcaptionbox{PT}
{
    \vspace{-2mm}
	\begin{tikzpicture}[scale=0.95]
		\begin{axis}[
		height=\columnwidth/2.5,
		width=\columnwidth/2.0,
		ylabel={\em Accuracy},
		xmin=1, xmax=9,
		xtick={1,2,3,4,5,6,7,8,9},
		xticklabels={1,3,5,10,20,40,60,80,100},
		every axis y label/.style={at={(current axis.north west)},right=3mm,above=0mm},
		every axis x label/.style={at={(current axis.south)},above=-7mm},
		label style={font=\scriptsize},
		tick label style={font=\scriptsize},
		]
		
		\addplot[dashed,color=red]
		table[x=ratio, y=linear, col sep=comma] {data/scale_pt.tex};
		
		\addplot[line width=0.25mm,color=magenta,mark=o]
		table[x=ratio, y=dhtr, col sep=comma] {data/scale_pt.tex};
		
		\addplot[line width=0.25mm,color=cyan,mark=triangle]
		table[x=ratio, y=mtraj, col sep=comma] {data/scale_pt.tex};
		
		\addplot[line width=0.25mm,color=green,mark=star]
		table[x=ratio, y=rntraj, col sep=comma] {data/scale_pt.tex};
		
		\addplot[line width=0.25mm,color=orange,mark=diamond]
		table[x=ratio, y=teri, col sep=comma] {data/scale_pt.tex};
		
		\addplot[line width=0.25mm,color=blue,mark=+]
		table[x=ratio, y=stged, col sep=comma] {data/scale_pt.tex};
		
		\addplot[line width=0.25mm,color=red,mark=triangle*]
		table[x=ratio, y=ours, col sep=comma] {data/scale_pt.tex};
		
		\end{axis}
	\end{tikzpicture}
}%
& \subcaptionbox{XA}
{
    \vspace{-2mm}
	\begin{tikzpicture}[scale=0.95]
		\begin{axis}[
		height=\columnwidth/2.5,
		width=\columnwidth/2.0,
		ylabel={\em Accuracy},
		xmin=1, xmax=9,
		xtick={1,2,3,4,5,6,7,8,9},
		xticklabels={1,3,5,10,20,40,60,80,100},
		every axis y label/.style={at={(current axis.north west)},right=3mm,above=0mm},
		every axis x label/.style={at={(current axis.south)},above=-7mm},
		label style={font=\scriptsize},
		tick label style={font=\scriptsize},
		]
		
		\addplot[dashed,color=red]
		table[x=ratio, y=linear, col sep=comma] {data/scale_xa.tex};
		
		\addplot[line width=0.25mm,color=magenta,mark=o]
		table[x=ratio, y=dhtr, col sep=comma] {data/scale_xa.tex};
		
		\addplot[line width=0.25mm,color=cyan,mark=triangle]
		table[x=ratio, y=mtraj, col sep=comma] {data/scale_xa.tex};
		
		\addplot[line width=0.25mm,color=green,mark=star]
		table[x=ratio, y=rntraj, col sep=comma] {data/scale_xa.tex};
		
		\addplot[line width=0.25mm,color=orange,mark=diamond]
		table[x=ratio, y=teri, col sep=comma] {data/scale_xa.tex};
		
		\addplot[line width=0.25mm,color=blue,mark=+]
		table[x=ratio, y=stged, col sep=comma] {data/scale_xa.tex};
		
		\addplot[line width=0.25mm,color=red,mark=triangle*]
		table[x=ratio, y=ours, col sep=comma] {data/scale_xa.tex};
		
		\end{axis}
	\end{tikzpicture}
} \\
\subcaptionbox{BJ}
{
    \vspace{-2mm}
	\begin{tikzpicture}[scale=0.95]
		\begin{axis}[
		height=\columnwidth/2.5,
		width=\columnwidth/2.0,
		ylabel={\em Accuracy},
		xmin=1, xmax=9,
		xtick={1,2,3,4,5,6,7,8,9},
		xticklabels={1,3,5,10,20,40,60,80,100},
		every axis y label/.style={at={(current axis.north west)},right=3mm,above=0mm},
		every axis x label/.style={at={(current axis.south)},above=-7mm},
		label style={font=\scriptsize},
		tick label style={font=\scriptsize},
		]
		
		\addplot[dashed,color=red]
		table[x=ratio, y=linear, col sep=comma] {data/scale_bj.tex};
		
		\addplot[line width=0.25mm,color=magenta,mark=o]
		table[x=ratio, y=dhtr, col sep=comma] {data/scale_bj.tex};
		
		\addplot[line width=0.25mm,color=cyan,mark=triangle]
		table[x=ratio, y=mtraj, col sep=comma] {data/scale_bj.tex};
		
		\addplot[line width=0.25mm,color=green,mark=star]
		table[x=ratio, y=rntraj, col sep=comma] {data/scale_bj.tex};
		
		\addplot[line width=0.25mm,color=orange,mark=diamond]
		table[x=ratio, y=teri, col sep=comma] {data/scale_bj.tex};
		
		\addplot[line width=0.25mm,color=blue,mark=+]
		table[x=ratio, y=stged, col sep=comma] {data/scale_bj.tex};
		
		\addplot[line width=0.25mm,color=red,mark=triangle*]
		table[x=ratio, y=ours, col sep=comma] {data/scale_bj.tex};
		
		\end{axis}
	\end{tikzpicture}
}%
& \subcaptionbox{CD}
{
    \vspace{-2mm}
	\begin{tikzpicture}[scale=0.95]
		\begin{axis}[
		height=\columnwidth/2.5,
		width=\columnwidth/2.0,
		ylabel={\em Accuracy},
		xmin=1, xmax=9,
		xtick={1,2,3,4,5,6,7,8,9},
		xticklabels={1,3,5,10,20,40,60,80,100},
		every axis y label/.style={at={(current axis.north west)},right=3mm,above=0mm},
		every axis x label/.style={at={(current axis.south)},above=-7mm},
		label style={font=\scriptsize},
		tick label style={font=\scriptsize},
		]
		
		\addplot[dashed,color=red]
		table[x=ratio, y=linear, col sep=comma] {data/scale_cd.tex};
		
		\addplot[line width=0.25mm,color=magenta,mark=o]
		table[x=ratio, y=dhtr, col sep=comma] {data/scale_cd.tex};
		
		\addplot[line width=0.25mm,color=cyan,mark=triangle]
		table[x=ratio, y=mtraj, col sep=comma] {data/scale_cd.tex};
		
		\addplot[line width=0.25mm,color=green,mark=star]
		table[x=ratio, y=rntraj, col sep=comma] {data/scale_cd.tex};
		
		\addplot[line width=0.25mm,color=orange,mark=diamond]
		table[x=ratio, y=teri, col sep=comma] {data/scale_cd.tex};
		
		\addplot[line width=0.25mm,color=blue,mark=+]
		table[x=ratio, y=stged, col sep=comma] {data/scale_cd.tex};
		
		\addplot[line width=0.25mm,color=red,mark=triangle*]
		table[x=ratio, y=ours, col sep=comma] {data/scale_cd.tex};
		
		\end{axis}
	\end{tikzpicture}
} \\

\end{tabular}

\vspace{-2mm}
\end{small}
\caption{Trajectory Recovery When Varying Training Data Size}
\label{fig:exp-scale}
\vspace{-4mm}
\end{figure}
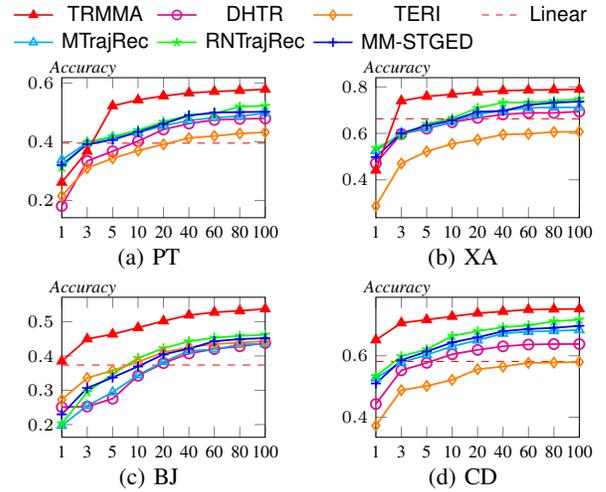

\header \textbf{Robustness v.s. Amount of Training Data.}
We evaluate the robustness of \algo and baselines using between 1\% and 100\% of the total training data, and report accuracy results in \fig~\ref{fig:exp-scale}. Note that \linear does not involve training and serves as a benchmark. Typically, reducing the amount of training data degrades a method's accuracy. The performance gap between \algo and the   competitors maintains under most settings. In \fig~\ref{fig:exp-scale}(c,d) for BJ and CD, \algo consistently outperforms existing methods under all settings. In \fig~\ref{fig:exp-scale}(a,b) for PT and XA, \algo achieves the best performance after using 3\% and 1\% of the training data, respectively.

\begin{table*}[!t]
    \small
    \centering
    \caption{Effectiveness of Map Matching. A larger value is better (in percentage). Best is in bold and runner-up is underlined.} 
    \vspace{-2mm}
    \label{tab:effect-mm}
    
    \renewcommand{\arraystretch}{1}
    \resizebox{0.96\textwidth}{!}{%
    \setlength{\tabcolsep}{2pt}
        \begin{tabular}{l|cccc|cccc|cccc|cccc}
            \hline
            & \multicolumn{4}{c|}{PT}& \multicolumn{4}{c|}{XA}     & \multicolumn{4}{c|}{BJ}     & \multicolumn{4}{c}{CD}       \\
            Methods & Precision & Recall & F1 & Jaccard    & Precision & Recall & F1 & Jaccard    & Precision & Recall & F1 & Jaccard    & Precision & Recall & F1 & Jaccard\\
            \hline
            {\nn} & {80.42} & {85.42} & {82.42} & {74.55}    & {79.01} & {89.79} & {82.69} & {75.03}    & {66.81} & {71.86} & {68.20} & {59.93}    & {72.29} & {87.24} & {77.32} & {69.10} \\
            \fmm & 86.34 & 83.71 & 84.74 & 78.08    & 93.60 & 91.85 & 92.49 & 88.84    & 72.51 & 70.36 & 70.69 & 63.82    & 89.14 & 88.39 & 88.34 & 83.94 \\
            \lhmm & 89.80 & 87.06 & 88.20 & 82.37 & \uline{95.53} & 94.14 & 94.62 & 91.84 & 75.30 & 72.35 & 73.08 & 65.34 & 91.19 & 90.69 & 90.57 & 87.10 \\
            \rntrajrec & 89.70 & 89.46 & 89.10 & 84.29 & 93.15 & 94.10 & 93.03 & 89.73 & \uline{78.82} & 76.64 & 76.80 & \uline{70.30} & 89.46 & 91.17 & 89.45 & 85.48 \\
            \deepmm & \uline{91.34} & \uline{90.95} & \uline{90.88} & \uline{86.22} & 95.40 & \uline{95.14} & \uline{95.06} & \uline{92.23} & 78.29 & \uline{77.66} & \uline{76.99} & 69.41 & \uline{94.99} & \uline{94.67} & \uline{94.58} & \uline{91.54} \\
            \graphmm & 87.01 & 88.84 & 87.26 & 79.13 & 92.84 & 94.62 & 92.75 & 87.06 & 75.39 & 73.84 & 72.32 & 62.82 & 88.53 & 92.56 & 89.31 & 82.23 \\
            \gts & \textbf{94.46} & \textbf{94.53} & \textbf{94.35} & \textbf{91.53}    & \textbf{97.20} & \textbf{97.97} & \textbf{97.36} & \textbf{95.97}   & \textbf{82.17} & \textbf{81.08} & \textbf{80.92} & \textbf{75.28}     & \textbf{96.27} & \textbf{97.51} & \textbf{96.54} & \textbf{94.94} \\
            \hline
        \end{tabular}%
    }
    \vspace{-4mm}
\end{table*}

\begin{figure}[!t]
\centering
\begin{small}
\resizebox{0.8\columnwidth}{!}{%
\vspace{-1mm}
\begin{tikzpicture}[scale=0.9]
    \begin{customlegend}[
        legend entries={\gts(ours),\lhmm,\rntrajrec,\deepmm,\graphmm,\fmm},
        legend columns=3,
        area legend,
        legend style={at={(0.45,1.15)},anchor=north,draw=none,font=\small,column sep=0.15cm}]
        
        \addlegendimage{preaction={fill, magenta}, pattern=dots}
        \addlegendimage{preaction={fill, myyellow},pattern={horizontal lines}}
        \addlegendimage{preaction={fill, mycyan},pattern=crosshatch}
        \addlegendimage{preaction={fill, mycyan2}, pattern=grid}
        \addlegendimage{preaction={fill, mypink},pattern=crosshatch dots}
        \addlegendimage{preaction={fill, myorange}, pattern={north east lines}}
    \end{customlegend}
\end{tikzpicture}
}
\\[-\lineskip]
\vspace{-1mm}
\captionsetup{margin={-0.5cm,0cm}}

\begin{tikzpicture}[scale=0.92]
\begin{axis}[
    height=\columnwidth/2.6,
    width=\columnwidth/1,
    axis lines=left,
    ybar=1.5pt,
    bar width=0.15cm,
    enlarge x limits=true,
    xtick={0.04,0.3,0.6,0.87},
    xticklabels={PT,XA,BJ,CD},
    xmin=0, xmax=0.9,
    ymin=0.1, ymax=50,
    ytick={0.1,1,10,100,1000,10000},
    ymode=log,
    log origin y=infty,
    log basis y={10},
    xtick style={draw=none},
    every axis y label/.style={at={(current axis.north west)},right=10mm,above=0mm},
    legend style={at={(0.02,0.98)},anchor=north west,cells={anchor=west},font=\tiny}
    ]
        \addplot[preaction={fill, magenta}, pattern=dots] coordinates {(0.3,0.19)};
        \addplot[preaction={fill, mycyan2}, pattern=grid] coordinates {(0.3,5.67)};
        \addplot[preaction={fill, myyellow},pattern={horizontal lines}] coordinates {(0.3,7.73)};
        \addplot[preaction={fill, myorange}, pattern={north east lines}] coordinates {(0.3,7.77)};
        \addplot[preaction={fill, mypink},pattern=crosshatch dots] coordinates {(0.3,11.58)};
        \addplot[preaction={fill, mycyan},pattern=crosshatch]coordinates {(0.3,15.95)};
        \node[above,black] at (axis cs:-0.015,5){$\boldsymbol{\star}$};

        \addplot[preaction={fill, magenta}, pattern=dots] coordinates {(0.4,0.16)};
        \addplot[preaction={fill, mycyan2}, pattern=grid] coordinates {(0.4,4.93)};
        \addplot[preaction={fill, myorange}, pattern={north east lines}] coordinates {(0.4,6.11)};
        \addplot[preaction={fill, mycyan},pattern=crosshatch]coordinates {(0.4,10.12)};
        \addplot[preaction={fill, myyellow},pattern={horizontal lines}] coordinates {(0.4,10.86)};
        \addplot[preaction={fill, mypink},pattern=crosshatch dots] coordinates {(0.4,14.56)};
        \node[above,black] at (axis cs:0.263,5){$\boldsymbol{\star}$};

        \addplot[preaction={fill, magenta}, pattern=dots] coordinates {(0.5,0.12)};
        \addplot[preaction={fill, mypink},pattern=crosshatch dots] coordinates {(0.5,4.04)};
        \addplot[preaction={fill, myyellow},pattern={horizontal lines}] coordinates {(0.5,5.07)};
        \addplot[preaction={fill, mycyan2}, pattern=grid] coordinates {(0.5,6.21)};
        \addplot[preaction={fill, myorange}, pattern={north east lines}] coordinates {(0.5,7.42)};
        \addplot[preaction={fill, mycyan},pattern=crosshatch]coordinates {(0.5,7.91)};
        \node[above,black] at (axis cs:0.604,5){$\boldsymbol{\star}$};
        
        \addplot[preaction={fill, magenta}, pattern=dots] coordinates {(0.6,0.22)};
        \addplot[preaction={fill, mycyan2}, pattern=grid] coordinates {(0.6,4.99)};
        \addplot[preaction={fill, myorange}, pattern={north east lines}] coordinates {(0.6,9.65)};
        \addplot[preaction={fill, mycyan},pattern=crosshatch]coordinates {(0.6,11.70)};
        \addplot[preaction={fill, myyellow},pattern={horizontal lines}] coordinates {(0.6,13.99)};
        \addplot[preaction={fill, mypink},pattern=crosshatch dots] coordinates {(0.6,15.72)};
        \node[above,black] at (axis cs:0.822,5){$\boldsymbol{\star}$};

\end{axis}
\end{tikzpicture}\vspace{-2mm}\hspace{2mm}\label{fig:infer_mm_pt}%

\vspace{-1mm}
\end{small}
\caption{Inference Time of Map Matching per 1000 Trajectories (in seconds) ($\boldsymbol{\star}$ marks the best competitor in \tab \ref{tab:effect-mm}).} 
\label{fig:infer_mm}
\vspace{-3mm}
\end{figure}
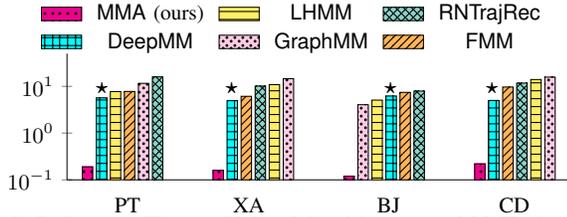

\begin{figure}[!t]
\centering
\begin{small}
\resizebox{0.8\columnwidth}{!}{%
\begin{tikzpicture}[scale=0.92]
    \begin{customlegend}[
        legend entries={\gts (ours),\lhmm,\rntrajrec,\deepmm,\graphmm},
        legend columns=3,
        area legend,
        legend style={at={(0.45,1.15)},anchor=north,draw=none,font=\small,column sep=0.15cm}]
        
        \addlegendimage{preaction={fill, magenta}, pattern=dots}
        \addlegendimage{preaction={fill, myyellow},pattern={horizontal lines}}
        \addlegendimage{preaction={fill, mycyan},pattern=crosshatch}
        \addlegendimage{preaction={fill, mycyan2}, pattern=grid}
        \addlegendimage{preaction={fill, mypink},pattern=crosshatch dots}
    \end{customlegend}
\end{tikzpicture}
}
\\[-\lineskip]
\vspace{-2mm}
\captionsetup{margin={-0.5cm,0cm}}

\begin{tikzpicture}[scale=0.92]
\begin{axis}[
    height=\columnwidth/2.4,
    width=\columnwidth/1,
    axis lines=left,
    ybar=1.5pt,
    bar width=0.15cm,
    xtick={0.1,0.35,0.55,0.8},
    xticklabels={PT,XA,BJ,CD},
    xmin=0, xmax=0.9,
    ymin=1, ymax=1000,
    ytick={0.1,1,10,100,1000,10000},
    ymode=log,
    log origin y=infty,
    log basis y={10},
    xtick style={draw=none},
    every axis y label/.style={at={(current axis.north west)},right=10mm,above=0mm},
    every axis x label/.style={at={(current axis.south)},above=-7mm},
    legend style={at={(0.02,0.98)},anchor=north west,cells={anchor=west},font=\tiny}
    ]
        \addplot[preaction={fill, magenta}, pattern=dots] coordinates {(0.3,152.5/60)};
        \addplot[preaction={fill, mycyan2}, pattern=grid] coordinates {(0.3,535.5/60)};
        \addplot[preaction={fill, myyellow},pattern={horizontal lines}] coordinates {(0.3,885.2/60)};
        \addplot[preaction={fill, mypink},pattern=crosshatch dots] coordinates {(0.3,1074.9/60)};
        \addplot[preaction={fill, mycyan},pattern=crosshatch]coordinates {(0.3,6582.1/60)};
        \node[above,black] at (axis cs:0.087,8){$\boldsymbol{\star}$};

        \addplot[preaction={fill, magenta}, pattern=dots] coordinates {(0.4,211.7/60)};
        \addplot[preaction={fill, mycyan2}, pattern=grid] coordinates {(0.4,621.8/60)};
        \addplot[preaction={fill, mypink},pattern=crosshatch dots] coordinates {(0.4,1853.7/60)};
        \addplot[preaction={fill, myyellow},pattern={horizontal lines}] coordinates {(0.4,2226.1/60)};
        \addplot[preaction={fill, mycyan},pattern=crosshatch]coordinates {(0.4,9177.5/60)};
        \node[above,black] at (axis cs:0.312,9){$\boldsymbol{\star}$};

        \addplot[preaction={fill, magenta}, pattern=dots] coordinates {(0.5,136.1/60)};
        \addplot[preaction={fill, mycyan2}, pattern=grid] coordinates {(0.5,712.7/60)};
        \addplot[preaction={fill, myyellow},pattern={horizontal lines}] coordinates {(0.5,788.2/60)};
        \addplot[preaction={fill, mypink},pattern=crosshatch dots] coordinates {(0.5,2361.6/60)};
        \addplot[preaction={fill, mycyan},pattern=crosshatch]coordinates {(0.5,53345.6/60)};
        \node[above,black] at (axis cs:0.537,10){$\boldsymbol{\star}$};

        \addplot[preaction={fill, magenta}, pattern=dots] coordinates {(0.6,349.6/60)};
        \addplot[preaction={fill, mycyan2}, pattern=grid] coordinates {(0.6,792.5/60)};
        \addplot[preaction={fill, myyellow},pattern={horizontal lines}] coordinates {(0.6,2744.1/60)};
        \addplot[preaction={fill, mypink},pattern=crosshatch dots] coordinates {(0.6,3827.4/60)};
        \addplot[preaction={fill, mycyan},pattern=crosshatch]coordinates {(0.6,13491.4/60)};
        \node[above,black] at (axis cs:0.762,12){$\boldsymbol{\star}$};

\end{axis}
\end{tikzpicture}\vspace{-2mm}\hspace{2mm}\label{fig:train_mm_pt}%

\vspace{-1mm}
\end{small}
\caption{Training Time per Epoch (in minutes) of Map Matching ($\boldsymbol{\star}$ marks the best competitor   in \tab \ref{tab:effect-mm}).} 
\label{fig:train_mm}
\vspace{-4mm}
\end{figure}
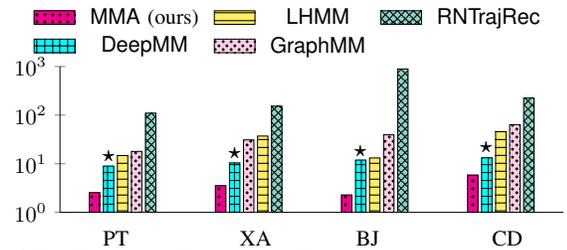

\subsection{Effectiveness and Efficiency of Map Matching}\label{sec:effectivenessandEfficiencyMM}

\noindent
\textbf{Effectiveness.}
\tab~\ref{tab:effect-mm} presents the map matching results of our method, \gts, compared to competitors. For trajectory $\tgps$, the metric values are based on segments in the returned route  $\Route$, not the recovered 
$\thigh$ in trajectory recovery above. 
As shown in \tab~\ref{tab:effect-mm}, \gts outperforms all methods across all four datasets under all metrics. For instance, on PT, \gts achieves a 91.53\% Jaccard score, surpassing \deepmm's 86.22\% by 5.31\%. On BJ, \gts has a recall of 81.08\%, 3.42\% higher than \deepmm's 77.66\%. This superior performance validates the effectiveness of our analysis and formulation in Section~\ref{sec:gtsconcepts}, treating it as a classification task over a small candidate segment set to identify the segment of a GPS point, and demonstrates the power of the proposed point and candidate segment embedding techniques in Section~\ref{sec:gtsArchi}.

\header
\textbf{Efficiency.} 
\fig~\ref{fig:infer_mm} reports the average inference time for map matching 1000 trajectories. Our method, \gts, is significantly faster than all other methods, often by orders of magnitude. For example, on the PT dataset, \gts requires only 0.19 seconds, whereas \deepmm requires 5.67 seconds, indicating a 29-fold speedup. \fig~\ref{fig:train_mm} reports the training time per epoch, except for \fmm, which does not require training. Similarly, \gts is faster to train than existing methods. The efficiency of \gts in both inference and training demonstrates the power of the proposed techniques in Section~\ref{sec:gps2seg}. Specifically, given a GPS point, we focus on a small candidate segment set to identify its segment in Section~\ref{sec:gtsconcepts}, and then develop efficient and effective embedding techniques in Section~\ref{sec:gtsArchi} to capture intrinsic patterns for map matching.

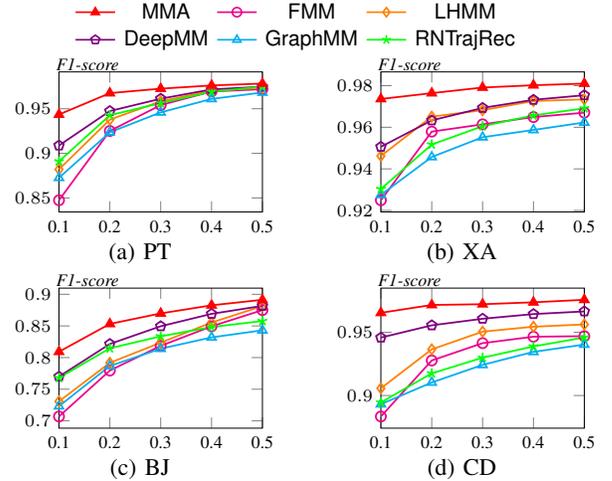
\begin{figure}[!t]
\centering
\begin{small}
\vspace{-1mm}
\resizebox{0.7\columnwidth}{!}{%
\begin{tikzpicture}[scale=0.85]
    \begin{customlegend}[legend columns=3,
	    legend entries={\gts,\fmm,\lhmm,\deepmm,\graphmm,\rntrajrec},
	    legend style={at={(0.45,1.15)},anchor=north,draw=none,font=\scriptsize,column sep=0.05cm}]
	    
	    \addlegendimage{line width=0.25mm,color=red,mark=triangle*}
 	    \addlegendimage{line width=0.25mm,color=magenta,mark=o}
 	    \addlegendimage{line width=0.25mm,color=orange,mark=diamond}
	    \addlegendimage{line width=0.25mm,color=violet,mark=pentagon}
	    \addlegendimage{line width=0.25mm,color=cyan,mark=triangle}
 	    \addlegendimage{line width=0.25mm,color=green,mark=star}
    \end{customlegend}
\end{tikzpicture}
}
\\[-\lineskip]
\vspace{-1mm}

\begin{tabular}{cc}

\subcaptionbox{PT}
{
    \vspace{-2mm}
	\begin{tikzpicture}[scale=0.95]
		\begin{axis}[
		height=\columnwidth/2.5,
		width=\columnwidth/2.0,
		ylabel={\em F1-score},
		xmin=1, xmax=5,
		xtick={1,2,3,4,5},
		xticklabels={0.1,0.2,0.3,0.4,0.5},
		every axis y label/.style={at={(current axis.north west)},right=4mm,above=0mm},
		every axis x label/.style={at={(current axis.south)},above=-7mm},
		label style={font=\scriptsize},
		tick label style={font=\scriptsize},
		]
		
		\addplot[line width=0.25mm,color=magenta,mark=o]
		table[x=ratio, y=fmm, col sep=comma] {data/mm_ratio_pt.tex};
		
		\addplot[line width=0.25mm,color=orange,mark=diamond]
		table[x=ratio, y=lhmm, col sep=comma] {data/mm_ratio_pt.tex};
		
		\addplot[line width=0.25mm,color=violet,mark=pentagon]
		table[x=ratio, y=deepmm, col sep=comma] {data/mm_ratio_pt.tex};
		
		\addplot[line width=0.25mm,color=cyan,mark=triangle]
		table[x=ratio, y=graphmm, col sep=comma] {data/mm_ratio_pt.tex};
		
		\addplot[line width=0.25mm,color=green,mark=star]
		table[x=ratio, y=rntraj, col sep=comma] {data/mm_ratio_pt.tex};
		
		\addplot[line width=0.25mm,color=red,mark=triangle*]
		table[x=ratio, y=ours, col sep=comma] {data/mm_ratio_pt.tex};
		
		\end{axis}
	\end{tikzpicture}
}%
& \subcaptionbox{XA}
{
    \vspace{-2mm}
	\begin{tikzpicture}[scale=0.95]
		\begin{axis}[
		height=\columnwidth/2.5,
		width=\columnwidth/2.0,
		ylabel={\em F1-score},
		xmin=1, xmax=5,
		xtick={1,2,3,4,5},
		xticklabels={0.1,0.2,0.3,0.4,0.5},
		every axis y label/.style={at={(current axis.north west)},right=4mm,above=0mm},
		every axis x label/.style={at={(current axis.south)},above=-7mm},
		label style={font=\scriptsize},
		tick label style={font=\scriptsize},
		]
		
		\addplot[line width=0.25mm,color=magenta,mark=o]
		table[x=ratio, y=fmm, col sep=comma] {data/mm_ratio_xa.tex};
		
		\addplot[line width=0.25mm,color=orange,mark=diamond]
		table[x=ratio, y=lhmm, col sep=comma] {data/mm_ratio_xa.tex};
		
		\addplot[line width=0.25mm,color=violet,mark=pentagon]
		table[x=ratio, y=deepmm, col sep=comma] {data/mm_ratio_xa.tex};
		
		\addplot[line width=0.25mm,color=cyan,mark=triangle]
		table[x=ratio, y=graphmm, col sep=comma] {data/mm_ratio_xa.tex};
		
		\addplot[line width=0.25mm,color=green,mark=star]
		table[x=ratio, y=rntraj, col sep=comma] {data/mm_ratio_xa.tex};
		
		\addplot[line width=0.25mm,color=red,mark=triangle*]
		table[x=ratio, y=ours, col sep=comma] {data/mm_ratio_xa.tex};
		
		\end{axis}
	\end{tikzpicture}
} \\
\subcaptionbox{BJ}
{
    \vspace{-2mm}
	\begin{tikzpicture}[scale=0.95]
		\begin{axis}[
		height=\columnwidth/2.5,
		width=\columnwidth/2.0,
		ylabel={\em F1-score},
		xmin=1, xmax=5,
		xtick={1,2,3,4,5},
		xticklabels={0.1,0.2,0.3,0.4,0.5},
		every axis y label/.style={at={(current axis.north west)},right=4mm,above=0mm},
		every axis x label/.style={at={(current axis.south)},above=-7mm},
		label style={font=\scriptsize},
		tick label style={font=\scriptsize},
		]
		
		\addplot[line width=0.25mm,color=magenta,mark=o]
		table[x=ratio, y=fmm, col sep=comma] {data/mm_ratio_bj.tex};
		
		\addplot[line width=0.25mm,color=orange,mark=diamond]
		table[x=ratio, y=lhmm, col sep=comma] {data/mm_ratio_bj.tex};
		
		\addplot[line width=0.25mm,color=violet,mark=pentagon]
		table[x=ratio, y=deepmm, col sep=comma] {data/mm_ratio_bj.tex};
		
		\addplot[line width=0.25mm,color=cyan,mark=triangle]
		table[x=ratio, y=graphmm, col sep=comma] {data/mm_ratio_bj.tex};
		
		\addplot[line width=0.25mm,color=green,mark=star]
		table[x=ratio, y=rntraj, col sep=comma] {data/mm_ratio_bj.tex};
		
		\addplot[line width=0.25mm,color=red,mark=triangle*]
		table[x=ratio, y=ours, col sep=comma] {data/mm_ratio_bj.tex};
		
		\end{axis}
	\end{tikzpicture}
}%
& \subcaptionbox{CD}
{
    \vspace{-2mm}
	\begin{tikzpicture}[scale=0.95]
		\begin{axis}[
		height=\columnwidth/2.5,
		width=\columnwidth/2.0,
		ylabel={\em F1-score},
		xmin=1, xmax=5,
		xtick={1,2,3,4,5},
		xticklabels={0.1,0.2,0.3,0.4,0.5},
		every axis y label/.style={at={(current axis.north west)},right=4mm,above=0mm},
		every axis x label/.style={at={(current axis.south)},above=-7mm},
		label style={font=\scriptsize},
		tick label style={font=\scriptsize},
		]
		
		\addplot[line width=0.25mm,color=magenta,mark=o]
		table[x=ratio, y=fmm, col sep=comma] {data/mm_ratio_cd.tex};
		
		\addplot[line width=0.25mm,color=orange,mark=diamond]
		table[x=ratio, y=lhmm, col sep=comma] {data/mm_ratio_cd.tex};
		
		\addplot[line width=0.25mm,color=violet,mark=pentagon]
		table[x=ratio, y=deepmm, col sep=comma] {data/mm_ratio_cd.tex};
		
		\addplot[line width=0.25mm,color=cyan,mark=triangle]
		table[x=ratio, y=graphmm, col sep=comma] {data/mm_ratio_cd.tex};
		
		\addplot[line width=0.25mm,color=green,mark=star]
		table[x=ratio, y=rntraj, col sep=comma] {data/mm_ratio_cd.tex};
		
		\addplot[line width=0.25mm,color=red,mark=triangle*]
		table[x=ratio, y=ours, col sep=comma] {data/mm_ratio_cd.tex};
		
		\end{axis}
	\end{tikzpicture}
} \\
\end{tabular}%

\vspace{-2mm}
\end{small}
\caption{Map Matching  with Varied Levels of Sparsity} 
\label{fig:exp-ratio-mm}
\vspace{-4mm}
\end{figure}

\header \textbf{Map Matching with Varied Levels of Sparsity.}
As mentioned, given a trajectory with the target $\eps$ sampling rate, we generate its sparse trajectory by random sampling with  time interval $\eps/\samp$, where $\samp$ controls the level of sparsity. We vary $\samp$ from 0.1 to 0.5, and report the F1 scores of map matching in \fig~\ref{fig:exp-ratio-mm}.
As trajectories become sparser with a smaller $\samp$, all methods achieve degraded performance. Nevertheless, \gts is the best across all sparsity levels on all datasets, demonstrating the effectiveness of our techniques.

\section{Conclusion}
We present \algo and \gts, efficient methods for accurate trajectory recovery and map matching, respectively, with \gts serving as the initial step for \algo. \gts formulates a classification task to map a GPS point to road segment within a small candidate set, and generates effective embeddings for accurate map matching. \algo then focuses on the segments of the route identified by \gts for recovery, avoiding the need to evaluate all road segments. It employs a dual-transformer encoding process to capture latent patterns and a decoding technique to predict position ratios and road segments of missing points. Extensive experiments on large real-world datasets demonstrate that \algo and \gts outperform existing methods in both result quality and efficiency.

\section*{Acknowledgment}

The work described in this paper was supported by grants from the Research Grants Council of Hong Kong Special Administrative Region, China (No. PolyU 25201221, PolyU 15205224, PolyU 152043/23E).
Jieming Shi is supported by NSFC No. 62202404, Otto Poon Charitable Foundation Smart Cities Research Institute (SCRI) P0051036-P0050643.
This work is supported by
Tencent Technology Co., Ltd. P0048511.

\clearpage

\bibliographystyle{IEEEtran}
\bibliography{main}

\begin{thebibliography}{10}
\providecommand{\url}[1]{#1}
\csname url@samestyle\endcsname
\providecommand{\newblock}{\relax}
\providecommand{\bibinfo}[2]{#2}
\providecommand{\BIBentrySTDinterwordspacing}{\spaceskip=0pt\relax}
\providecommand{\BIBentryALTinterwordstretchfactor}{4}
\providecommand{\BIBentryALTinterwordspacing}{\spaceskip=\fontdimen2\font plus
\BIBentryALTinterwordstretchfactor\fontdimen3\font minus \fontdimen4\font\relax}
\providecommand{\BIBforeignlanguage}[2]{{%
\expandafter\ifx\csname l@#1\endcsname\relax
\typeout{** WARNING: IEEEtran.bst: No hyphenation pattern has been}%
\typeout{** loaded for the language `#1'. Using the pattern for}%
\typeout{** the default language instead.}%
\else
\language=\csname l@#1\endcsname
\fi
#2}}
\providecommand{\BIBdecl}{\relax}
\BIBdecl

\bibitem{NMLR21}
J.~Jain, V.~Bagadia, S.~Manchanda, and S.~Ranu, ``Neuromlr: Robust {\&} reliable route recommendation on road networks,'' in \emph{NeurIPS}, 2021, pp. 22\,070--22\,082.

\bibitem{DRPK23}
W.~Tian, J.~Shi, S.~Luo, H.~Li, X.~Xie, and Y.~Zou, ``Effective and efficient route planning using historical trajectories on road networks,'' \emph{{PVLDB}}, vol.~16, no.~10, pp. 2512--2524, 2023.

\bibitem{DeepOD20}
H.~Yuan, G.~Li, Z.~Bao, and L.~Feng, ``Effective travel time estimation: When historical trajectories over road networks matter,'' in \emph{{SIGMOD}}, 2020, pp. 2135--2149.

\bibitem{Yuan00020B22}
H.~Yuan, G.~Li, and Z.~Bao, ``Route travel time estimation on {A} road network revisited: Heterogeneity, proximity, periodicity and dynamicity,'' \emph{{PVLDB}}, vol.~16, no.~3, pp. 393--405, 2022.

\bibitem{VRE2022}
H.~Lan, J.~Xie, Z.~Bao, F.~Li, W.~Tian, F.~Wang, S.~Wang, and A.~Zhang, ``{VRE:} {A} versatile, robust, and economical trajectory data system,'' \emph{PVLDB}, vol.~15, no.~12, pp. 3398--3410, 2022.

\bibitem{FangG0XGJ23}
Z.~Fang, S.~Gong, L.~Chen, J.~Xu, Y.~Gao, and C.~S. Jensen, ``Ghost: {A} general framework for high-performance online similarity queries over distributed trajectory streams,'' \emph{{SIGMOD}}, vol.~1, no.~2, pp. 173:1--173:25, 2023.

\bibitem{Jin000023}
J.~Jin, P.~Cheng, L.~Chen, X.~Lin, and W.~Zhang, ``Efficient non-learning similar subtrajectory search,'' \emph{PVLDB}, vol.~16, no.~11, pp. 3111--3123, 2023.

\bibitem{FangPCDG21}
Z.~Fang, L.~Pan, L.~Chen, Y.~Du, and Y.~Gao, ``{MDTP:} {A} multi-source deep traffic prediction framework over spatio-temporal trajectory data,'' \emph{PVLDB}, vol.~14, no.~8, pp. 1289--1297, 2021.

\bibitem{YuanCL24}
H.~Yuan, G.~Cong, and G.~Li, ``Nuhuo: An effective estimation model for traffic speed histogram imputation on {A} road network,'' \emph{PVLDB}, vol.~17, no.~7, pp. 1605--1617, 2024.

\bibitem{RNTrajRec23}
Y.~Chen, H.~Zhang, W.~Sun, and B.~Zheng, ``Rntrajrec: Road network enhanced trajectory recovery with spatial-temporal transformer,'' in \emph{{ICDE}}, 2023, pp. 829--842.

\bibitem{LHMM24ICDE}
W.~Shi, J.~Xu, J.~Fang, P.~Chao, A.~Liu, and X.~Zhou, ``{LHMM:} {A} learning enhanced {HMM} model for cellular trajectory map matching,'' in \emph{{ICDE}}, 2023, pp. 2429--2442.

\bibitem{mm10}
J.~Yuan, Y.~Zheng, C.~Zhang, X.~Xie, and G.~Sun, ``An interactive-voting based map matching algorithm,'' in \emph{{MDM}}, 2010, pp. 43--52.

\bibitem{GraphMM24TKDE}
Y.~Liu, Q.~Ge, W.~Luo, Q.~Huang, L.~Zou, H.~Wang, X.~Li, and C.~Liu, ``Graphmm: Graph-based vehicular map matching by leveraging trajectory and road correlations,'' \emph{{IEEE} Trans. Knowl. Data Eng.}, vol.~36, no.~1, pp. 184--198, 2024.

\bibitem{MTrajRec21}
H.~Ren, S.~Ruan, Y.~Li, J.~Bao, C.~Meng, R.~Li, and Y.~Zheng, ``Mtrajrec: Map-constrained trajectory recovery via seq2seq multi-task learning,'' in \emph{{KDD}}, 2021, pp. 1410--1419.

\bibitem{ShiTZ0WY24}
D.~Shi, Y.~Tong, Z.~Zhou, K.~Xu, Z.~Wang, and J.~Ye, ``Graph-constrained diffusion for end-to-end path planning,'' in \emph{{ICLR}}, 2024.

\bibitem{PinkH08}
O.~Pink and B.~Hummel, ``A statistical approach to map matching using road network geometry, topology and vehicular motion constraints,'' in \emph{{ITSC}}, 2008, pp. 862--867.

\bibitem{HMM09}
P.~Newson and J.~Krumm, ``Hidden markov map matching through noise and sparseness,'' in \emph{{SIGSPATIAL}}, 2009, pp. 336--343.

\bibitem{DMM24TKDE}
Z.~Shen, K.~Yang, X.~Zhao, J.~Zou, W.~Du, and J.~Wu, ``Dmm: A deep reinforcement learning based map matching framework for cellular data,'' \emph{{IEEE} Trans. Knowl. Data Eng.}, vol.~36, no.~10, pp. 5120--5137, 2024.

\bibitem{STGED24}
T.~Wei, Y.~Lin, Y.~Lin, S.~Guo, L.~Zhang, and H.~Wan, ``Micro-macro spatial-temporal graph-based encoder-decoder for map-constrained trajectory recovery,'' \emph{{IEEE} Trans. Knowl. Data Eng.}, vol.~36, no.~11, pp. 6574--6587, 2024.

\bibitem{DHTR21}
J.~Wang, N.~Wu, X.~Lu, W.~X. Zhao, and K.~Feng, ``Deep trajectory recovery with fine-grained calibration using kalman filter,'' \emph{{IEEE} Trans. Knowl. Data Eng.}, vol.~33, no.~3, pp. 921--934, 2021.

\bibitem{TERI23}
Y.~Chen, G.~Cong, and C.~Anda, ``{TERI:} an effective framework for trajectory recovery with irregular time intervals,'' \emph{{PVLDB}}, vol.~17, no.~3, pp. 414--426, 2023.

\bibitem{MuslehM23}
M.~Musleh and M.~F. Mokbel, ``A demonstration of {KAMEL:} {A} scalable bert-based system for trajectory imputation,'' in \emph{{SIGMOD}}, 2023, pp. 191--194.

\bibitem{ElshrifIM22}
M.~M. Elshrif, K.~Isufaj, and M.~F. Mokbel, ``Network-less trajectory imputation,'' in \emph{{SIGSPATIAL}}, 2022, pp. 8:1--8:10.

\bibitem{AttnMove21}
T.~Xia, Y.~Qi, J.~Feng, F.~Xu, F.~Sun, D.~Guo, and Y.~Li, ``Attnmove: History enhanced trajectory recovery via attentional network,'' in \emph{{AAAI}}, 2021, pp. 4494--4502.

\bibitem{PeriodicMove21}
H.~Sun, C.~Yang, L.~Deng, F.~Zhou, F.~Huang, and K.~Zheng, ``Periodicmove: Shift-aware human mobility recovery with graph neural network,'' in \emph{{CIKM}}, 2021, pp. 1734--1743.

\bibitem{XiZLGXH19}
D.~Xi, F.~Zhuang, Y.~Liu, J.~Gu, H.~Xiong, and Q.~He, ``Modelling of bi-directional spatio-temporal dependence and users' dynamic preferences for missing {POI} check-in identification,'' in \emph{{AAAI}}, 2019, pp. 5458--5465.

\bibitem{STMM09}
Y.~Lou, C.~Zhang, Y.~Zheng, X.~Xie, W.~Wang, and Y.~Huang, ``Map-matching for low-sampling-rate {GPS} trajectories,'' in \emph{{SIGSPATIAL}}, 2009, pp. 352--361.

\bibitem{fmm2018}
C.~Yang and G.~Gid{\'{o}}falvi, ``Fast map matching, an algorithm integrating hidden markov model with precomputation,'' \emph{Int. J. Geogr. Inf. Sci.}, vol.~32, no.~3, pp. 547--570, 2018.

\bibitem{HRIS12}
K.~Zheng, Y.~Zheng, X.~Xie, and X.~Zhou, ``Reducing uncertainty of low-sampling-rate trajectories,'' in \emph{ICDE}, 2012, pp. 1144--1155.

\bibitem{NMM14}
P.~Banerjee, S.~Ranu, and S.~Raghavan, ``Inferring uncertain trajectories from partial observations,'' in \emph{ICDM}, 2014, pp. 30--39.

\bibitem{AntMapper18}
Y.~Gong, E.~Chen, X.~Zhang, L.~M. Ni, and J.~Zhang, ``Antmapper: An ant colony-based map matching approach for trajectory-based applications,'' \emph{{IEEE} Trans. Intell. Transp. Syst.}, vol.~19, no.~2, pp. 390--401, 2018.

\bibitem{DeepMM22TMC}
J.~Feng, Y.~Li, K.~Zhao, Z.~Xu, T.~Xia, J.~Zhang, and D.~Jin, ``Deepmm: Deep learning based map matching with data augmentation,'' \emph{{IEEE} Trans. Mob. Comput.}, vol.~21, no.~7, pp. 2372--2384, 2022.

\bibitem{L2MM23TKDD}
L.~Jiang, C.~Chen, and C.~Chen, ``{L2MM:} learning to map matching with deep models for low-quality {GPS} trajectory data,'' \emph{{ACM} Trans. Knowl. Discov. Data}, vol.~17, no.~3, pp. 39:1--39:25, 2023.

\bibitem{tvec18}
X.~Li, K.~Zhao, G.~Cong, C.~S. Jensen, and W.~Wei, ``Deep representation learning for trajectory similarity computation,'' in \emph{{ICDE}}, 2018, pp. 617--628.

\bibitem{NeuTraj19}
D.~Yao, G.~Cong, C.~Zhang, and J.~Bi, ``Computing trajectory similarity in linear time: {A} generic seed-guided neural metric learning approach,'' in \emph{{ICDE}}, 2019, pp. 1358--1369.

\bibitem{T3S21}
P.~Yang, H.~Wang, Y.~Zhang, L.~Qin, W.~Zhang, and X.~Lin, ``{T3S:} effective representation learning for trajectory similarity computation,'' in \emph{{ICDE}}, 2021, pp. 2183--2188.

\bibitem{TrajGAT22}
D.~Yao, H.~Hu, L.~Du, G.~Cong, S.~Han, and J.~Bi, ``Trajgat: {A} graph-based long-term dependency modeling approach for trajectory similarity computation,'' in \emph{{KDD}}, 2022, pp. 2275--2285.

\bibitem{TrajCL23}
Y.~Chang, J.~Qi, Y.~Liang, and E.~Tanin, ``Contrastive trajectory similarity learning with dual-feature attention,'' in \emph{{ICDE}}, 2023, pp. 2933--2945.

\bibitem{STVec22}
Z.~Fang, Y.~Du, X.~Zhu, D.~Hu, L.~Chen, Y.~Gao, and C.~S. Jensen, ``Spatio-temporal trajectory similarity learning in road networks,'' in \emph{{KDD}}, 2022, pp. 347--356.

\bibitem{GTS21}
P.~Han, J.~Wang, D.~Yao, S.~Shang, and X.~Zhang, ``A graph-based approach for trajectory similarity computation in spatial networks,'' in \emph{{KDD}}, 2021, pp. 556--564.

\bibitem{renfro2021analysis}
\BIBentryALTinterwordspacing
B.~A. Renfro, M.~Stein, E.~B. Reed, and A.~Finn, ``An analysis of global positioning system standard positioning service performance for 2022,'' \emph{Space and Geophysics Laboratory Applied Research Laboratories, The University of Texas at Austin}. [Online]. Available: \url{https://www.gps.gov/systems/gps/performance/}
\BIBentrySTDinterwordspacing

\bibitem{rtree97}
S.~T. Leutenegger, J.~M. Edgington, and M.~A. L{\'{o}}pez, ``{STR:} {A} simple and efficient algorithm for r-tree packing,'' in \emph{{ICDE}}, 1997, pp. 497--506.

\bibitem{node2vec16}
A.~Grover and J.~Leskovec, ``node2vec: Scalable feature learning for networks,'' in \emph{KDD}, 2016, pp. 855--864.

\bibitem{relu11}
X.~Glorot, A.~Bordes, and Y.~Bengio, ``Deep sparse rectifier neural networks,'' in \emph{{AISTATS}}, 2011, pp. 315--323.

\bibitem{trans17}
A.~Vaswani, N.~Shazeer, N.~Parmar, J.~Uszkoreit, L.~Jones, A.~N. Gomez, L.~Kaiser, and I.~Polosukhin, ``Attention is all you need,'' in \emph{{NIPS}}, 2017, pp. 5998--6008.

\bibitem{GRU14}
K.~Cho, B.~van Merrienboer, D.~Bahdanau, and Y.~Bengio, ``On the properties of neural machine translation: Encoder-decoder approaches,'' in \emph{EMNLP}, 2014, pp. 103--111.

\bibitem{Porto}
``Porto dataset,'' \url{https://www.kaggle.com/c/pkdd-15-predict-taxi-service-trajectory-i/data}.

\bibitem{DITA18}
Z.~Shang, G.~Li, and Z.~Bao, ``{DITA:} distributed in-memory trajectory analytics,'' in \emph{{SIGMOD}}, 2018, pp. 725--740.

\bibitem{OpenStreetMap}
``Openstreetmap,'' \url{https://www.openstreetmap.org/}.

\end{thebibliography}

\end{document}